\documentclass{svjour3}               
\smartqed  							
\usepackage{graphicx}
\usepackage{fix-cm}
\usepackage{amsmath}
\usepackage{bbding} 
\newcommand{\etal}{\textit{et al. }}
\newcommand{\ie}  {i.e. }
\newcommand{\eg}  {e.g. }
\newcommand{\fig} {Fig. }

\newcommand{\sect} {Section }
\newcommand{\cf} {cf}

\journalname{Journal of Low Temperature Physics}

\begin{document} 
 
\title{Textures of Superfluid $^3$He-B in Applied Flow and Comparison with Hydrostatic Theory}

\author{R. de Graaf \and V.B. Eltsov \and J.J. Hosio \and P.J. Heikkinen \and M.Krusius  }

\institute{R. de Graaf (\Envelope) \and V.B. Eltsov \and J.J. Hosio \and P.J. Heikkinen \and M. Krusius \newline
	Low Temperature Laboratory, Aalto University, School of Science and Technology, P.O. Box 15100, 00076 AALTO, Espoo, Finland \newline
	\email{rdegraaf@ltl.hut.fi}
}
 
\date{Received: date / Accepted: date} 

\maketitle
    
\begin{abstract} 
	Measurements of the order parameter texture of rotating superfluid $^3$He-B have been performed as a function of the applied azimuthal counterflow velocity down to temperatures of $0.2\,T_{\rm c}$.
The results are compared to the hydrostatic theory of $^3$He-B. 
Good agreement is found at all measured temperatures and rotation velocities when the flow anisotropy contribution to the textural free energy is adjusted. This gives a superfluid energy gap $\Delta(T)$ which agrees with that measured by Todoshchenko \textit{et al.}, with $\Delta(0)=1.97\, k_{\rm B} T_{\rm c}$ at 29.0 bar. 
The B-phase susceptibility, longitudinal resonance frequency, and textural phase transition have been extracted from the measurements as a function of temperature and azimuthal counterflow velocity.  Owing to decreasing absorption intensities the present measuring method, based on the line shape analysis of the NMR spectrum, loses its sensitivity with decreasing temperature. However, we find that in practice the measurement of vortex numbers and counterflow velocities is still feasible down to $0.2\,T_{\rm c}$
\end{abstract}

\section{Introduction}
\label{introduction}

Non-invasive nuclear magnetic resonance (NMR) measurements on superfluid $^3$He-B in a rotating cylindrical environment provide information on the axially symmetric order parameter texture. This method - based on an analysis of the line shape of the NMR spectrum - has been the foremost means of measuring the number of rectilinear vortices in a stationary state of rotation \cite{kopu_2000}.

A prominent feature in the NMR line shape is the so-called \textit{counterflow} (henceforth \cf) peak, which appears when there is sufficient flow velocity in the azimuthal direction. 
Counterflow is the difference between the velocities $v_{\rm n}$ of the normal component and that of the superfluid component $v_{\rm s}$ in the presence of $N$ axially aligned vortices, \ie $v_{\rm cf} = v_{\rm n} - v_{\rm s}$. In stationary state, rectilinear vortices form a cluster along the axis of rotation with a distribution which corresponds to the equilibrium vortex number density of the applied rotation. The radius of the cluster can be smaller than that of the system if $N<N_{\rm v}$, where $N_{\rm v}\approx 2\pi R^2\Omega_{\rm v}/ \kappa$ is the equilibrium number of vortices. Here $\kappa$ is the quantum of circulation, $R$ the radius of the system and $\Omega_{\rm v}$ the rotation velocity associated with $N_{\rm v}$ vortices in equilibrium state. For an example of a vortex cluster configuration, see \fig 1 in \cite{hanninen_2009}. Inside the vortex cluster the average superfluid velocity $\langle{v}_{\rm s}\rangle \approx v_{\rm n} = \Omega r$, while outside the centrally located cluster $v_{\rm s} = \kappa N / 2\pi r$.  The local cf velocity $v_{\rm cf}$ is thus approximately zero inside the cluster, while outside
\begin{equation}
	\vec{v}_{\rm cf}  = \vec{\Omega}\times \vec{r} - \Omega_{\rm v}R (R/r) \hat{e}_\phi = \left[\Omega - \Omega_{\rm v} \left(\frac{R}{r}\right)^2\right] r\,\hat{e}_\phi {\rm .}
\end{equation}

In contrast, flow in the axial direction is expressed by an increase of absorption in the spectrum near the Larmor peak. With \textit{continuous wave} (cw) NMR, the sweep time of the whole spectrum can be significantly longer than the characteristic time of the dynamics, in which case one has to resort to sweeping that part of the spectrum which expresses the vortex dynamics best: absorption of the \cf-peak or absorption near the Larmor frequency.  In this paper we only consider states where transient effects have ceased and the vortex configuration with rectilinear vortices is in the stationary state.

\begin{figure}[hp]
	\begin{minipage}{170pt}
		\begin{picture}(170,415)
			\put(20,360)  { \includegraphics[width=0.70\textwidth]{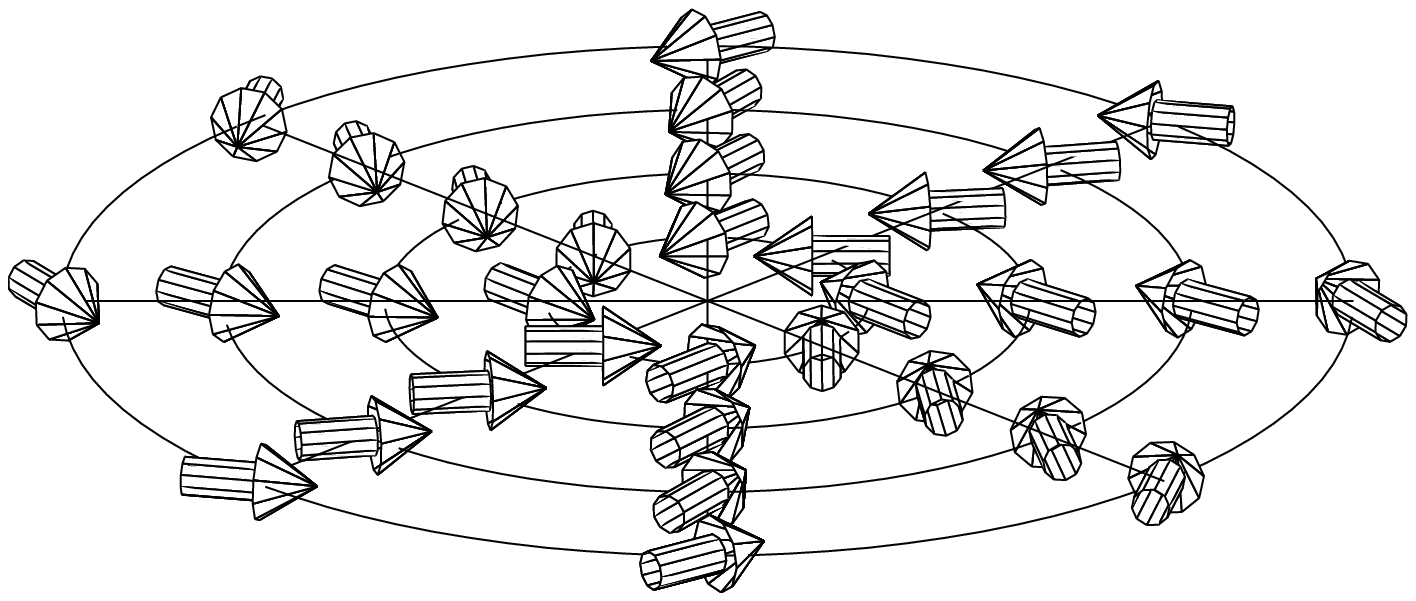} }
			\put(20,360)  { $\alpha$ }
			\put(20,305)  { \includegraphics[width=0.70\textwidth]{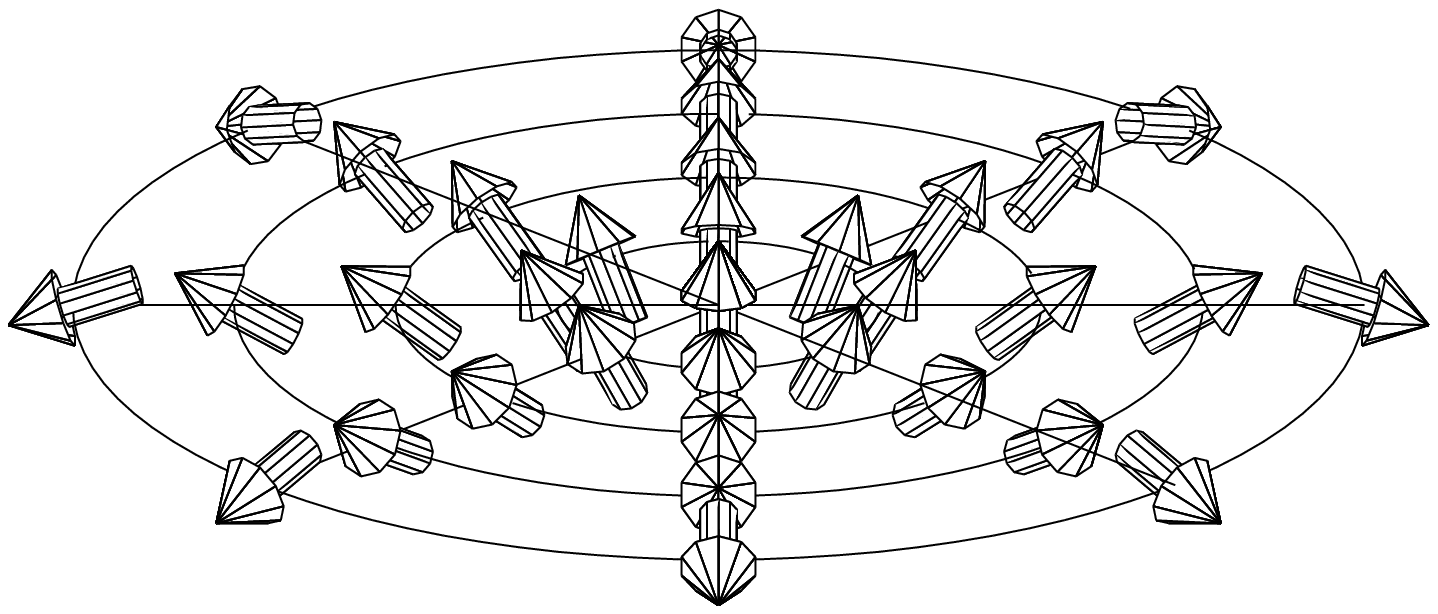} }
			\put(20,305)  { $\beta$ }
			\put(155,280) { \includegraphics[width=1.00\textwidth]{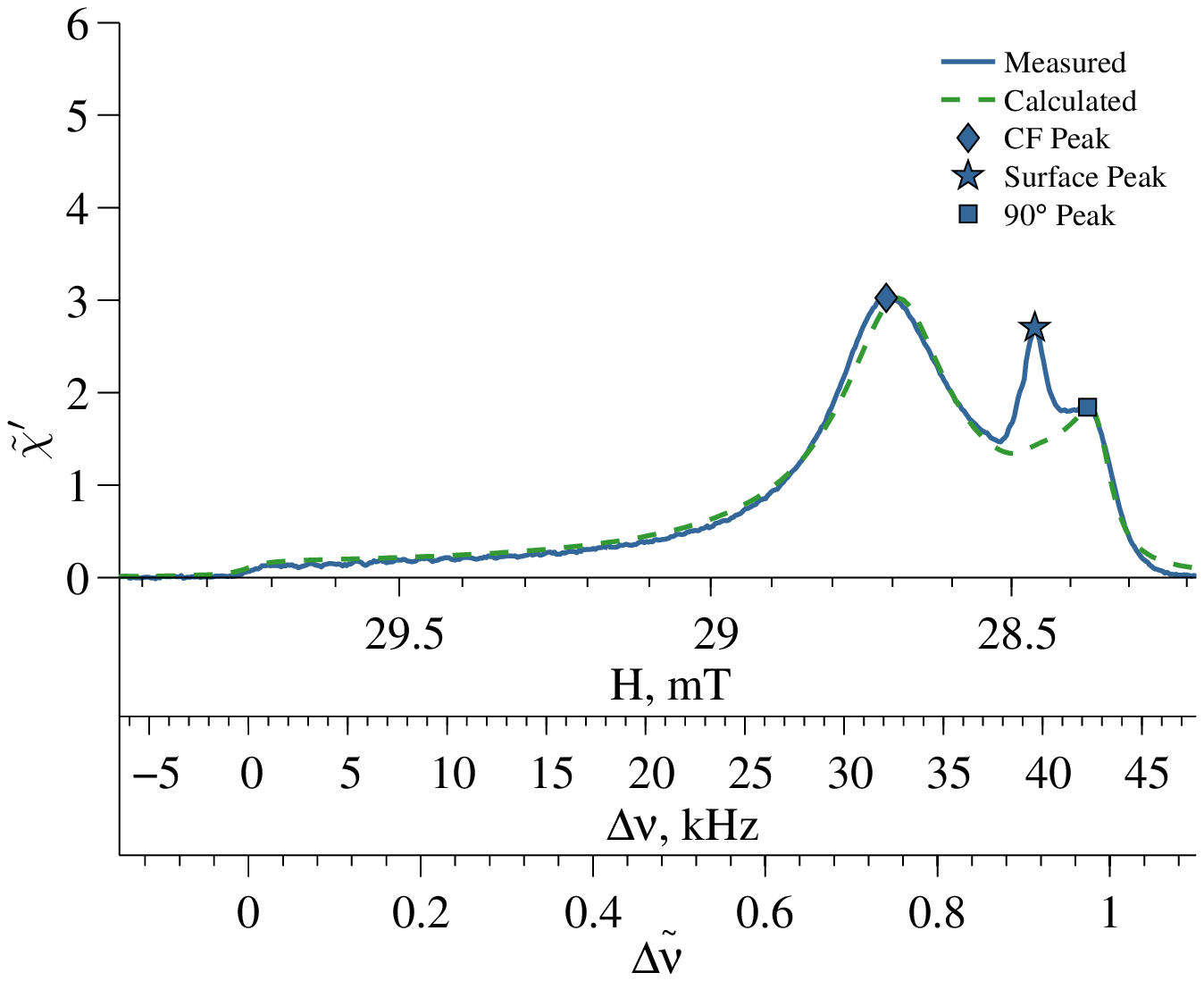} }
			\put(175,353) { \includegraphics[width=0.42\textwidth]{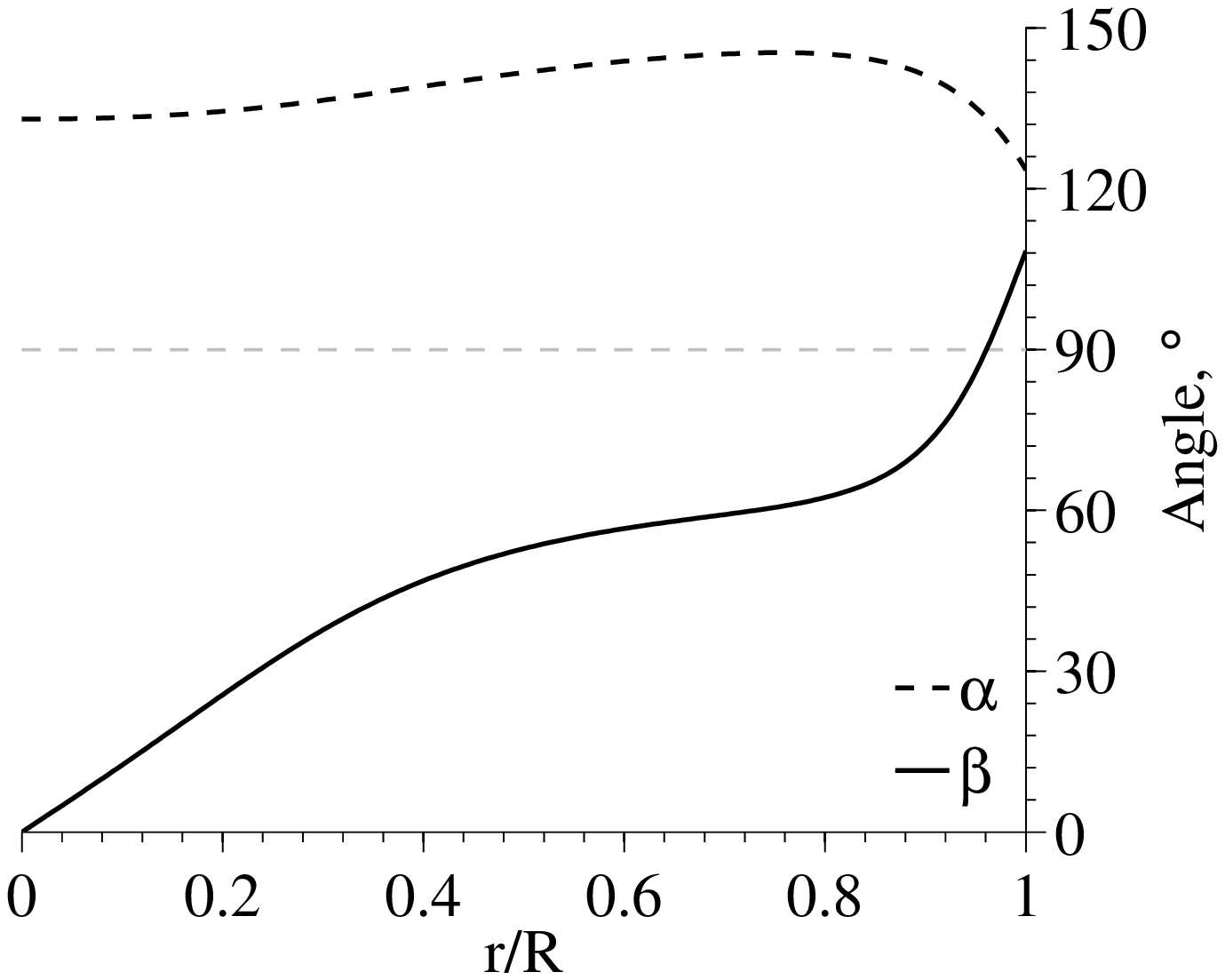} }
			\put(0,315)   { \rotatebox{90}{\textbf{Extended Flare Out}} } 
			\put(20,220)  { \includegraphics[width=0.70\textwidth]{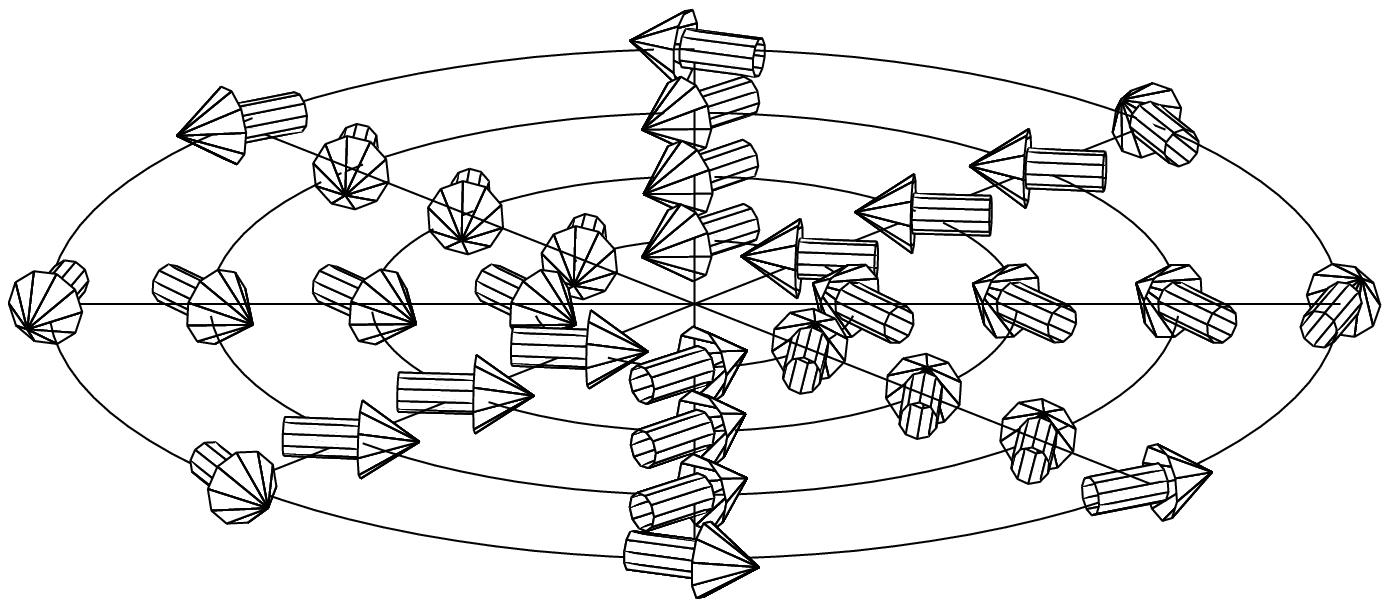} }
			\put(20,220)  { $\alpha$ }
			\put(20,165)  { \includegraphics[width=0.70\textwidth]{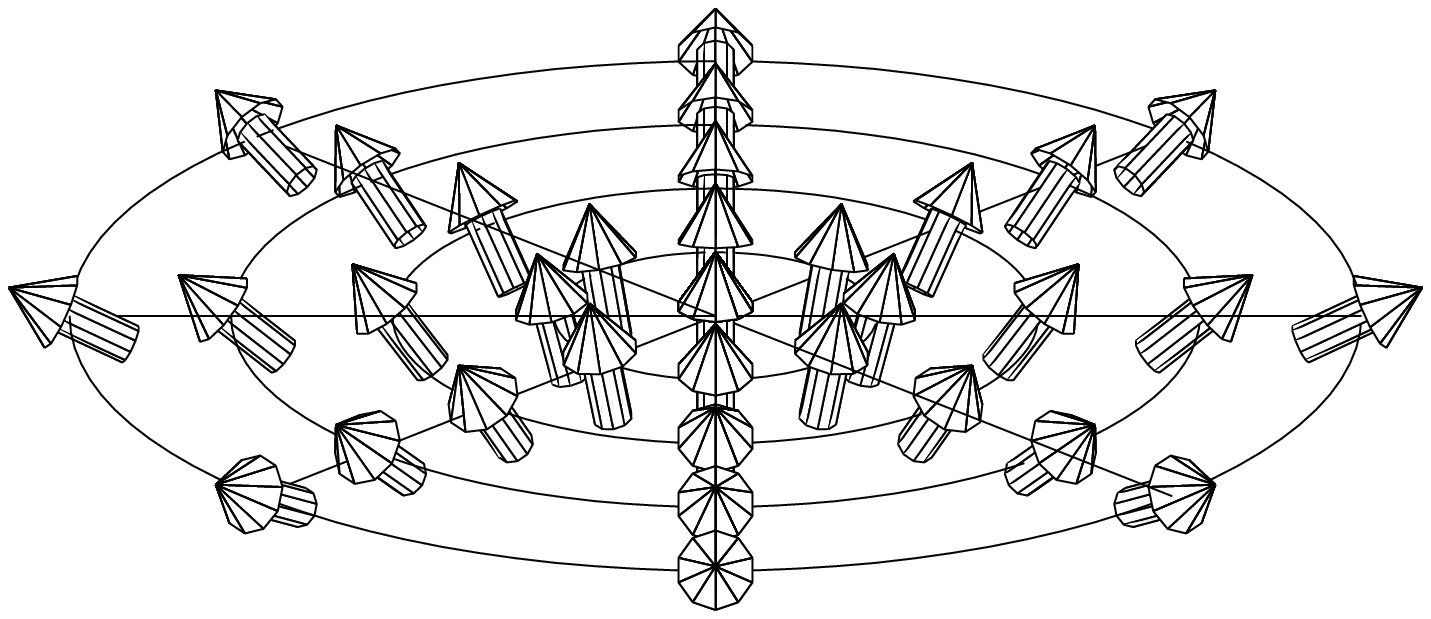} }
			\put(20,165)  { $\beta$ }
			\put(155,140) { \includegraphics[width=1.00\textwidth]{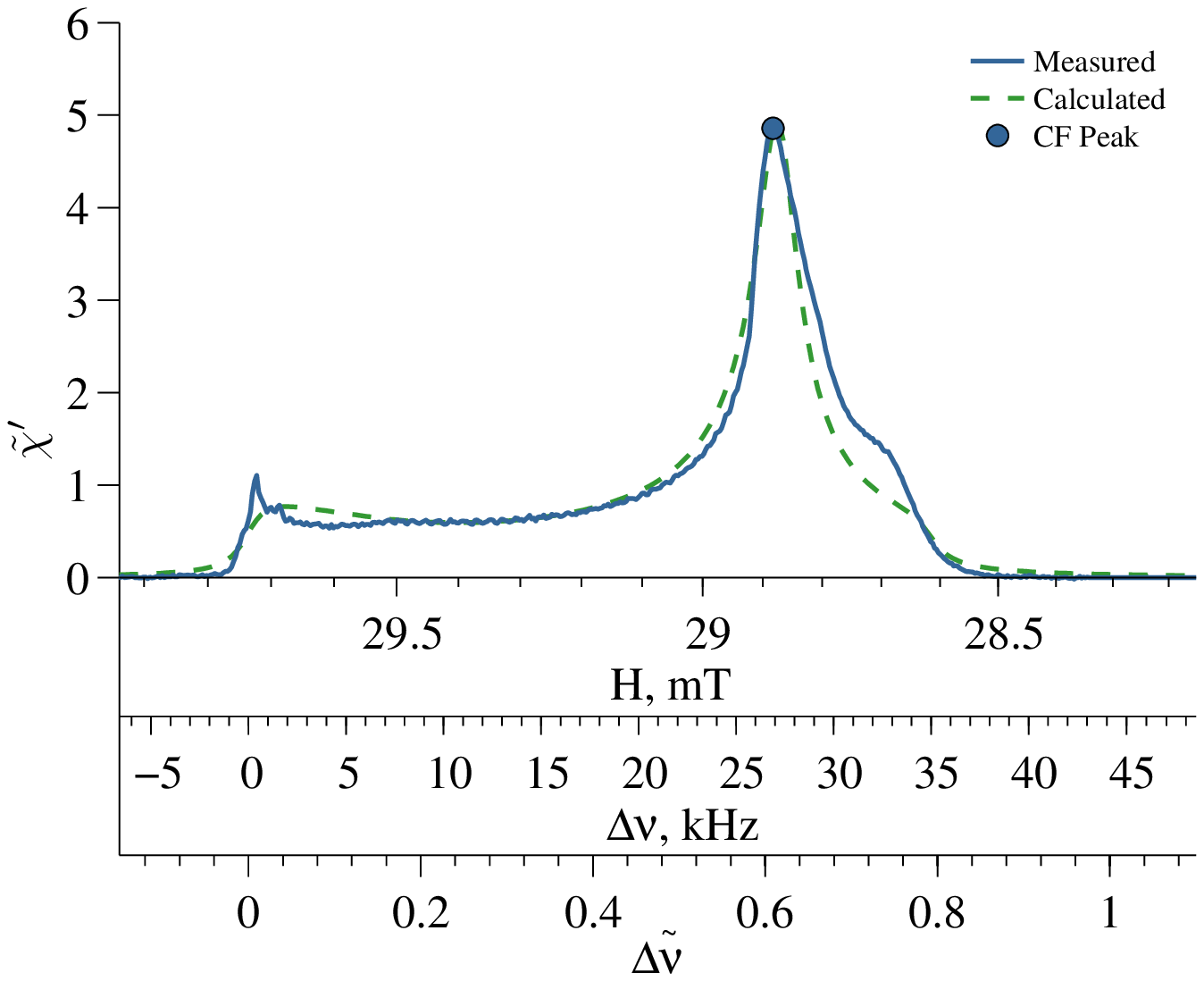} }
			\put(175,218) { \includegraphics[width=0.42\textwidth]{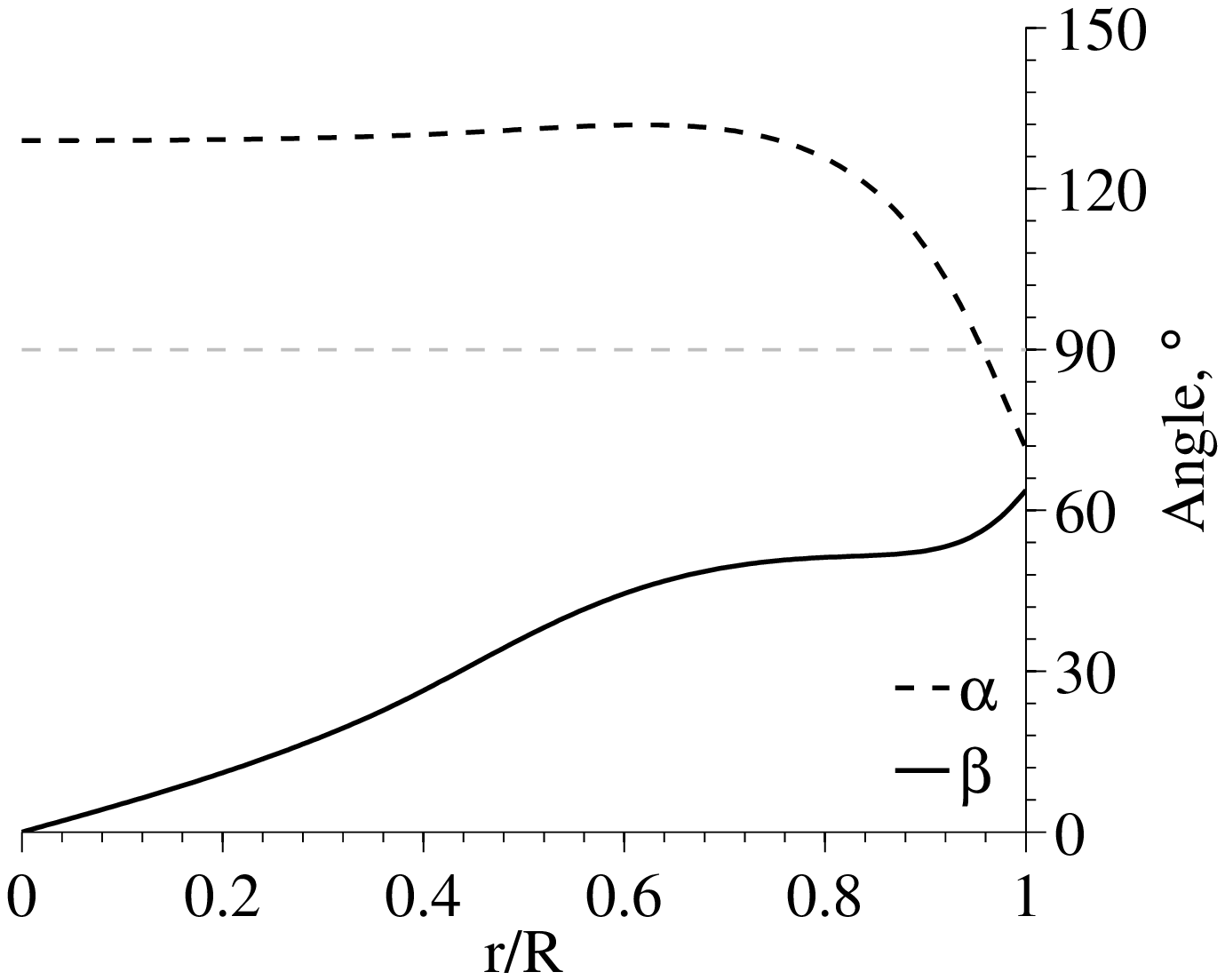} }			
			\put(0,180)   { \rotatebox{90}{\textbf{Parted Flare Out}} } 
			\put(20,80)   { \includegraphics[width=0.70\textwidth]{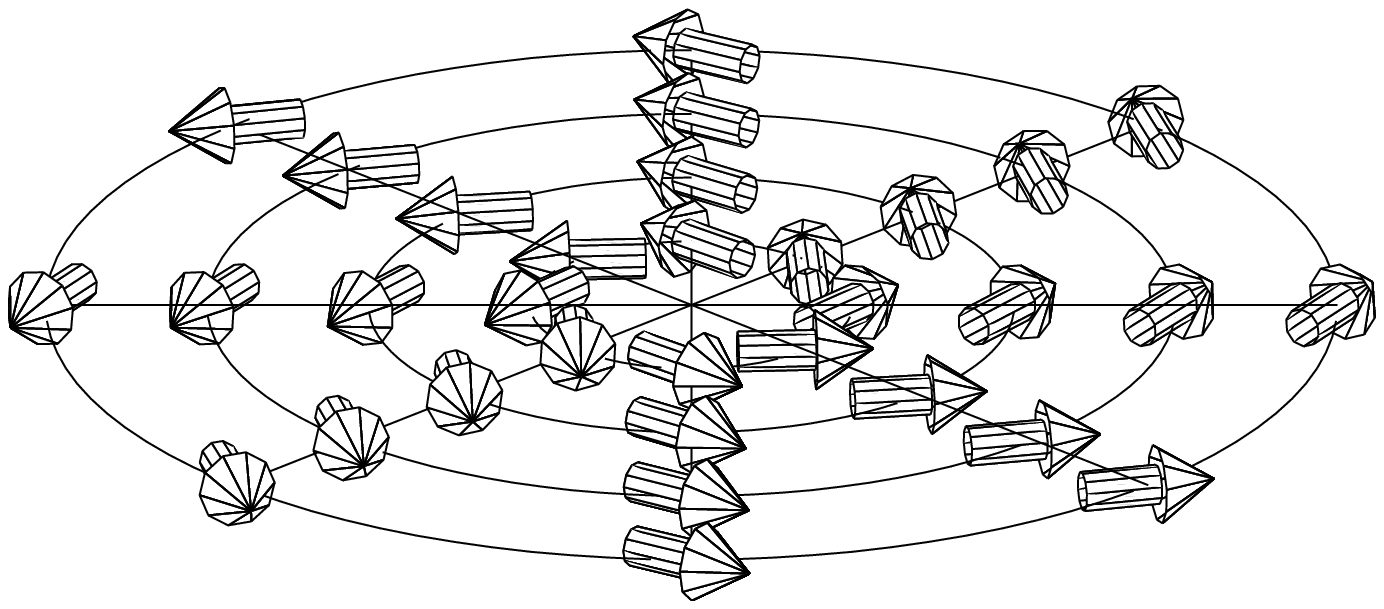} }
			\put(20,80)   { $\alpha$ }
			\put(20,25)   { \includegraphics[width=0.70\textwidth]{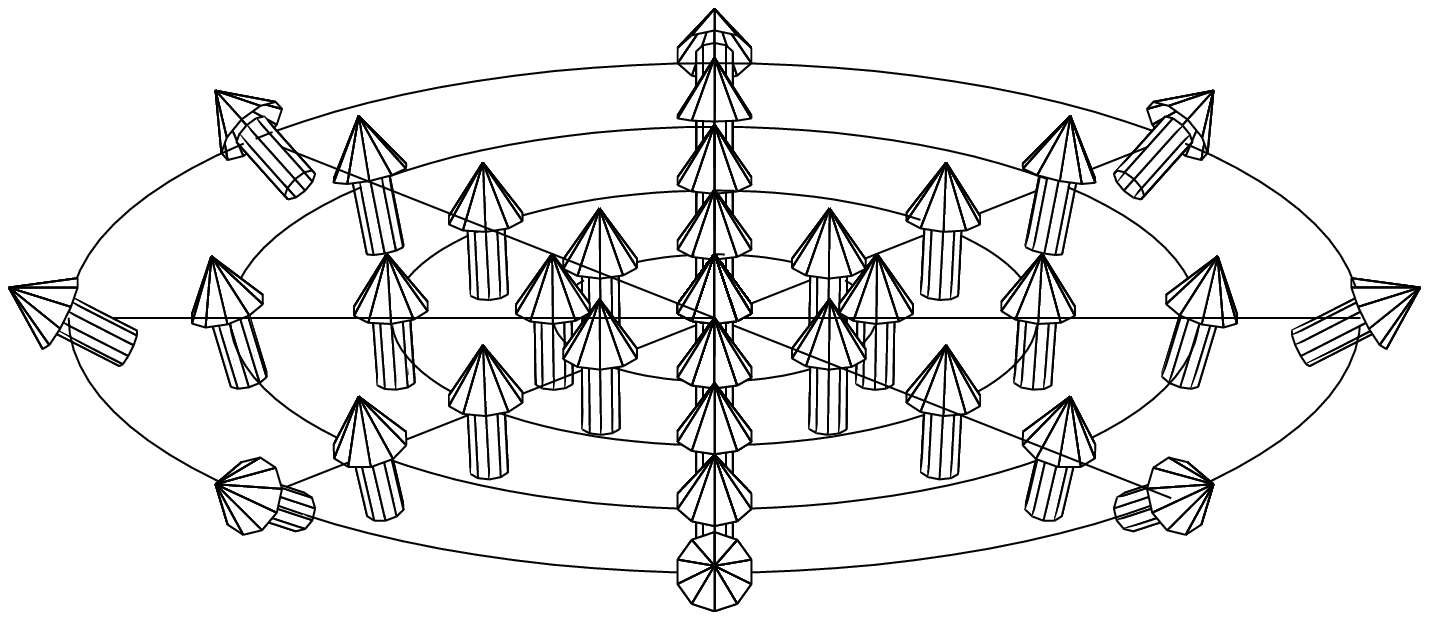} }
			\put(20,25)   { $\beta$ }
			\put(155,0)   { \includegraphics[width=1.00\textwidth]{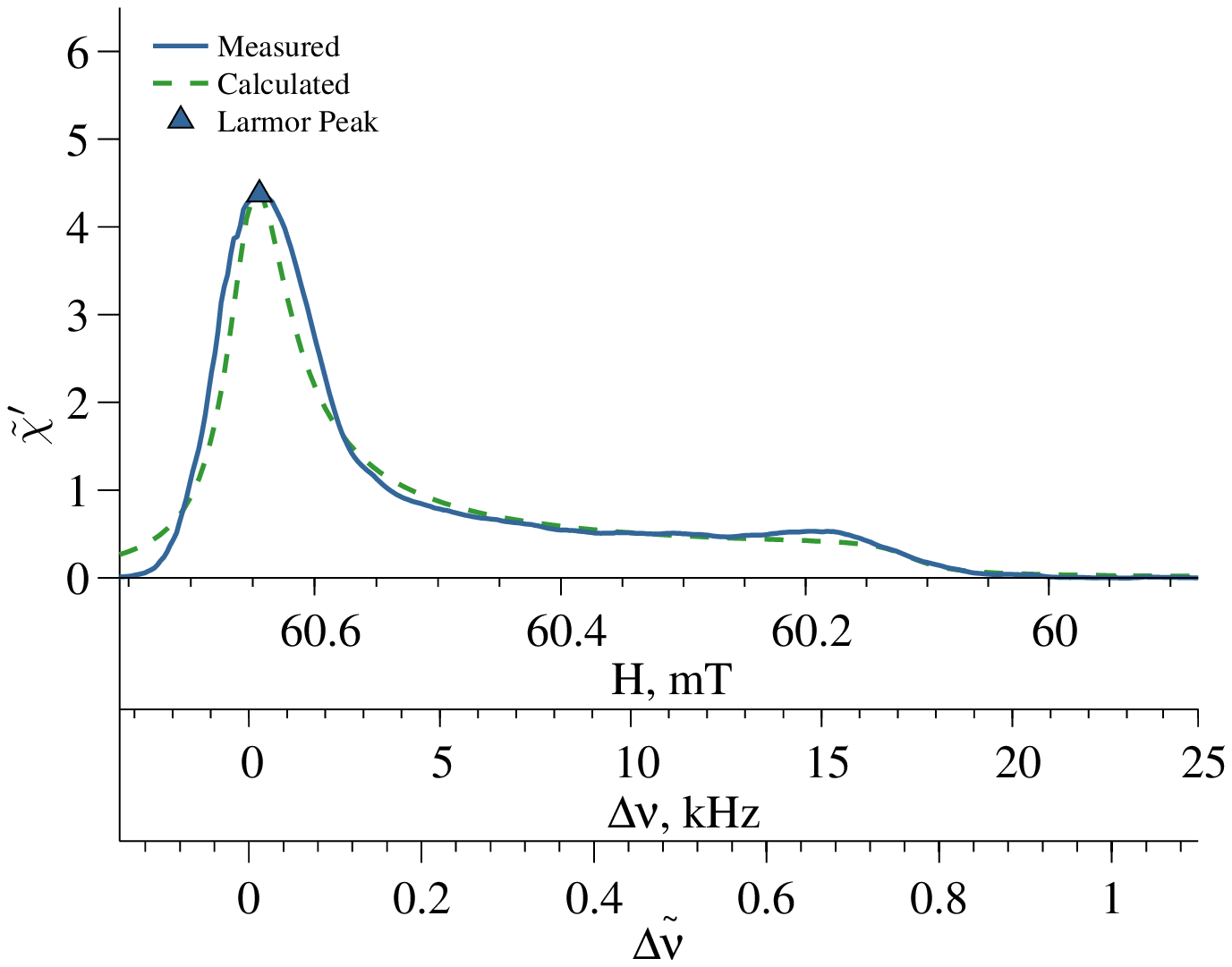} }
			\put(235,68)  { \includegraphics[width=0.50\textwidth]{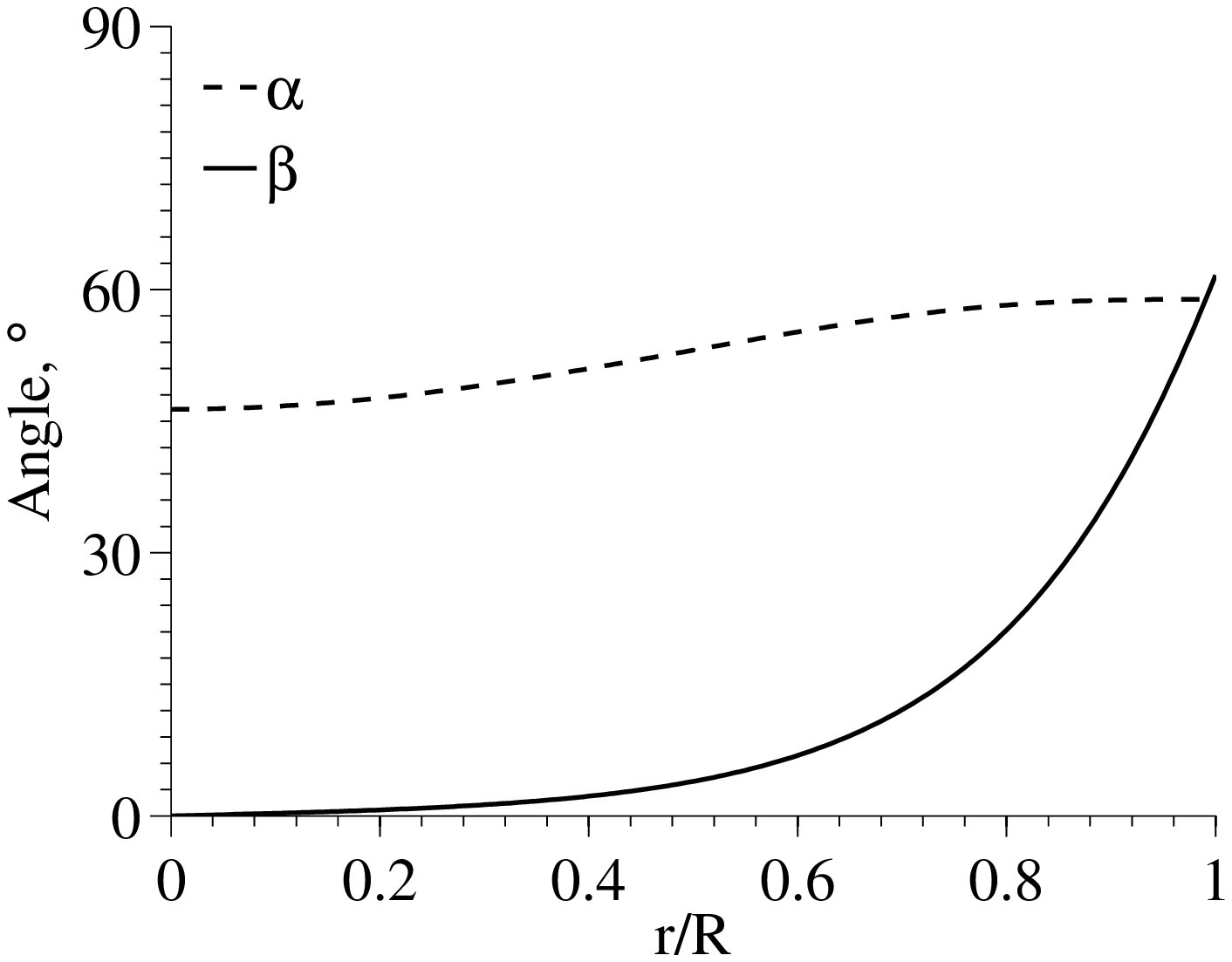} }
			\put(0,40)    { \rotatebox{90}{\textbf{Simple Flare Out}} } 
		\end{picture}
	\end{minipage}
	\caption{\textbf{Order parameter textures}. The measured and calculated NMR line shapes of the \textit{simple}, \textit{parted} and \textit{extended} flare out textures of $^3$He-B in a cylindrical cell with radius $R=3\,$mm. The measured spectra are presented as a function (horizontal axis) of the magnetic field $H$, or equivalently the frequency shift $\Delta\nu = \nu -\nu_{\rm L}$ from the Larmor frequency $\nu_{\rm L}$, or as a reduced frequency shift $\Delta\tilde{\nu}=\sin^2\beta$. Here the line shapes are normalized to the latter axis. The associated textures, \ie $\hat{\vec{n}}$-vector representation (by means of the angles $\alpha$ and $\beta$) as a function of reduced radius are depicted in the inset and on the left. Parameters of the \textit{simple} flare out texture: fork temperature $T=0.25\,T_{\rm c}$, pressure $p=29\,$bar, rotation velocity $\Omega=0.2\,$rad/s with no vortices, cw-NMR frequency $\nu_{\rm rf}=1.967\,$MHz and Larmor field $H_{\rm L}=60.65\,{\rm mT}$; \textit{parted flare out} texture: $T=0.31\,T_{\rm c}$, $\Omega=0.9\,$rad/s with vortex cluster $\Omega_{\rm v}=0.1\,$rad/s, $\nu_{\rm rf}=965.0\,$kHz and $H_{\rm L}=29.75\,{\rm mT}$; \textit{extended flare out} texture: $T=0.36\,T_{\rm c}$, $\Omega=0.7\,$rad/s with no vortices, $\nu_{\rm rf}=965.0\,$kHz and $H_{\rm L}=29.75\,{\rm mT}$. In the calculations, temperature $T$ (controlling the temperature dependent frequency shift $\Omega_B$) was a fitting parameter. Other fitting parameters; \textit{simple} flare out texture: field velocity parameter $\lambda_{\rm HV}=1.4\,{\rm kg/m^3T^2}$ and field inhomogeneity $\Delta H/H=8.3\cdot10^{-4}$; \textit{parted} flare out texture: $\lambda_{\rm HV}=3.1\,{\rm kg/m^3T^2}$ and $\Delta H/H=8.8\cdot10^{-4}$; \textit{extended} flare out texture: $\lambda_{\rm HV}=5.9\,{\rm kg/m^3T^2}$ and $\Delta H/H=8.8\cdot10^{-4}$. 
	 }
	\label{fig:textures and nmr line shapes}
\end{figure}

Examples of NMR spectra, measured in different conditions of flow at $0.2\div0.3\,T_{\rm c}$ temperatures, are shown in \fig \ref{fig:textures and nmr line shapes}. The spectra display absorption peaks which are shifted differently from the Larmor value. The shift and peak intensity of these absorption maxima identify the flow conditions. The integrated absorption intensity and simultaneously the shift of the peak at given rotation velocity $\Omega$ specify the number of vortices $N$ in the central vortex cluster and thereby the azimuthal counterflow circulating around the cluster. 

This NMR measurement technique based on the line shape analysis has been extensively used, \eg in measurements on dynamic remanent vortices \cite{solntsev_2007}, the onset temperature $T_{\rm on}$ to turbulence \cite{de_graaf_2008}, turbulent vortex front propagation \cite{eltsov_2007} and the stability of flow in the low temperature regime \cite{eltsov_2010}.
While in some experiments the measured spectra are used in a qualitative way (\eg the presence or absence of the \cf-peak), measurements where the number of rectilinear vortices in the sample is important need an (experimental) calibration for the number of vortices (or the azimuthal cf expressed as rotation velocity $\Omega_{\rm cf}=\Omega - \Omega_{\rm v}$) as a function of the cf peak height.

In this paper we compare the measured NMR response with the calculated spectra quantitatively. We start by reminding about the hydrostatic model of superfluid $^3$He-B in a cylindrical geometry \cite{thuneberg_2001,kopu_2007}. We discuss the superfluid order parameter textures, textural energies in applied flow and the relation to NMR. After the discussion of the experimental setup, we show results on measured texture transitions, on the susceptibility, and list spectra as a function of the azimuthal cf velocity. We compare measurements with calculations and obtain good agreement when we fit the density anisotropy, $\lambda_{\rm HV}$, of the superfluid. 
The discussion on the measured parameter $\lambda_{\rm HV}$ includes its relation to the energy gap $\Delta$ and considerations about its magnitude. We conclude the paper with the identification of a newly found resonance peak in the NMR spectrum. The peak is formed due to a non-local resonance in the potential well created by the texture near the cylindrical boundary of the experimental cell.

\section{Hydrostatic Theory of $^3$He-B}
In the B-phase\footnote[2]{The B-phase is also referred to as the BW-state after the discoverers R. Balian and N.R. Werthamer \cite{balian_1963}} of $^3$He, the Cooper pair with total spin $\vec{s}$ and angular momentum $\vec{l}$ is condensed to a state where the angle $\theta$ between \vec{s} and \vec{l} is fixed, due to interactions between the nuclear dipole moments of the helium atoms, to $\theta = \arccos(-\frac{1}{4}) \approx 104^\circ$ \cite{osheroff_1974}. In this configuration the total angular momentum $\vec{j}=0$, while $l=s=1$ \cite{balian_1963}. The order parameter matrix can be written as \cite{vollhardt_1990}
\begin{equation}
	A_{\mu j} = \Delta_{\mu \nu} R_{\nu j}(\hat{\vec{n}},\theta) e^{i\phi}{\rm ,}
\end{equation}
with $\Delta(T)$ the temperature dependent energy-gap, $\phi$ the condensate's phase, and $R_{ij}$ the rotation matrix describing the rotation of the spin and orbital coordinates relative to each other by $\theta$ around the axis oriented along the unit vector $\hat{\vec{n}}$:
\begin{equation}
	R_{ij}(\hat{\vec{n}},\theta) = \cos\theta \, \delta_{ij} + (1-\cos\theta)\hat{n}_i\hat{n}_j - \sin\theta \, \epsilon_{ijk} \, \hat{n}_k {\rm .}
\end{equation} 
Spherical coordinates are used for the unit vector $\hat{\vec{n}}$ with $\alpha$ and $\beta$ denoting the azimuthal and polar angles with respect to the axis of the static magnetic field $\vec{H}$. The orientational distribution of the $\hat{\vec{n}}$-vector over real space is called the \textit{texture}.
On length scales longer than the dipole-dipole interaction $\xi_{\rm D}\approx 10\,\mu {\rm m}$, a number of different interactions give rise to textural anisotropy \cite{hakonen_1989} in the bulk and at the surface. We start with four bulk terms and continue later with the surface term. The magnetic orientational free energy term
\begin{equation}
	F_{\rm DH} = -a \int d^3 r(\hat{\vec{n}}\cdot \vec{H})^2
	\label{eq: orientational free energy}
\end{equation}
results from the gap anisotropy in a magnetic field $H$. The dipole velocity term is
\begin{equation}
	F_{\rm DV} = -\lambda_{\rm DV} \int d^3 r [ \hat{\vec{n}}\cdot (\vec{v_s} - \vec{v_n})]^2 {\rm .}
\end{equation}
The gradient term
\begin{equation}
	F_{\rm G} = \int d^3r
	\left [
	 \lambda_{\rm G1} \frac{\partial R_{\alpha i}}{\partial r_i} \frac{\partial R_{\alpha j}}{\partial r_j}
	 +\lambda_{\rm G2} \frac{\partial R_{\alpha j}}{\partial r_i} \frac{\partial R_{\alpha j}}{\partial r_i}
	\right ]
\end{equation}
reflects the coherence of the superfluid state -- the stiffness of the order parameter with respect to spatial change -- and resists rapid spatial change in the texture.
The field velocity term
\begin{equation}
	F_{\rm HV} = -\lambda_{\rm HV} \int d^3 r [ \vec{H} \cdot \overleftrightarrow{R} \cdot (\vec{v_s} - \vec{v_n})]^2
	\label{eq: field velocity energy}
\end{equation}
arises from the anisotropy axis $\vec{l}$ preferring to orient in the direction of the flow, which changes the direction of the anisotropy axis of the energy gap $\Delta$.

We continue with the surface term. Here we assume that the curvature of the surface boundary is small and that the length scale of the distorted region is small compared to the dipole length $\xi_{\rm D}$. The surface gives rise to the surface field term
\begin{equation}
	F_{\rm SH} = - d \int_S d^2 r(\vec{H} \cdot \overleftrightarrow{R} \cdot \hat{\vec{s}})^2 {\rm }.
	\label{eq: surface field term}
\end{equation}
Here the unit vector $\hat{\vec{s}}$ is perpendicular to the surface and points toward the superfluid. For explicit expressions of the parameters $a$, $\lambda_{\rm DV}$, $\lambda_{\rm G1}$, $\lambda_{\rm G2}$, $d$ we refer to formulas (35)...(42) in \cite{thuneberg_2001}.
The field velocity parameter $\lambda_{\rm HV}$ is discussed here explicitly, since we use it later on as a fitting parameter, and to extract information on the energy gap $\Delta(T)$:
\begin{eqnarray}
	\lambda_{\rm HV}
		& = &	\frac{\rho}{\Delta^2}
				\frac{m^*/m}{(1+\frac{1}{3}F^{\rm s}_1Y)^2}
				\left [
					\frac{
						\frac{1}{2}\hbar\gamma\mu_0(1+\frac{1}{5}F^{\rm a}_2)
					}{
						1+F^{\rm a}_0(\frac{2}{3} + \frac{1}{3}Y) + \frac{1}{5}F^{\rm a}_2(\frac{1}{3} + (\frac{2}{3} + F^{\rm a}_0)Y)
					}
				\right ]^2 \nonumber \\
		& & 		\times
					\left [
						Z_3 - \frac{9}{10}Z_5 + \frac{9}{10}\frac{Z^2_5}{Z_3} - \frac{3}{2}Z_7
						+ \frac{3F^{\rm a}_2Z_3}{50(1+\frac{1}{5}F^{\rm a}_2)} (3Z_5 - 2Z_3)
					\right ] {\rm ,}
	\label{eq: field velocity parameter}
\end{eqnarray}
with $\rho$ the fluid density, $m^*$ the effective mass, $\gamma/2\pi=-32.435{\rm MHz/T}$ the gyromagnetic ratio,  $F^{\rm s}_l$ the symmetric and $F^{\rm a}_l$ the anti-symmetric Fermi-liquid parameter. The temperature-dependent functions $Z_j$ are defined as
\begin{eqnarray}
Z_j=\pi k_B T \Delta^{j-1} \sum_{n=-\infty}^{\infty}(\epsilon^2_n+\Delta^2)^{-j/2} {\rm ,}
\end{eqnarray} 
where the Matsubara energies are $\epsilon_n=\pi T (2n-1)$ with $n=0,\pm1, ..., \pm \infty$ and the Yosida function is given by $Y=1-Z_3(T)$. 

So the question arises: how do the bulk terms (\ref{eq: orientational free energy})...(\ref{eq: field velocity energy}) and the surface term (\ref{eq: surface field term}) affect the texture in a cylindrical environment? 
  
The rotational symmetry and continuity at $r<R$  means that at $r=0$: $\hat{\vec{n}} \parallel \vec{H}$, \ie $\beta=0$. At the boundary, the orbital momentum density $\vec{L}$ is perpendicular to the surface and $\sin^2\beta = 0.8$ (\ie $\beta\approx63.4^\circ$). In the intermediate region the texture is smoothed out by the gradient term $F_{\rm G}$. When the superfluid is stationary, \ie non-rotating, the \textit{simple} flare out texture is the lowest free energy state. See \fig \ref{fig:textures and nmr line shapes} for an example of the \textit{simple} flare out texture.

By rotating and applying azimuthal counterflow to the system, a negative energy contribution $F_{\rm HV}$ arises from the $\vec{l}$ vector orienting along the direction of the flow, which changes the direction of the density anisotropy, $\delta\rho^{(n)} = \rho^{(n)}_\parallel - \rho^{(n)}_\perp$, and reduces the total energy. This has a direct effect on the $\beta$ angle, see inset in \fig \ref{fig:textures and nmr line shapes} of the \textit{parted} flare out texture. The ends of the $\beta$ curve stay fixed at $r=0$ and $r=R$, but the intermediate slope moves to the left. At the critical velocity $\Omega_{\rm c1}$ the curve starts to show a plane near the surface boundary (where the cf is the largest). The angle $\beta$ on the surface is slightly moved due to the stress from the surface gradient energy term $F_{\rm SG}$. The resulting texture is now the \textit{parted} flare out texture with the cf-peak as a growing prominent feature.

At the critical velocity $\Omega_{\rm c2}$, the region with almost constant $\beta$ has grown to its maximum, but not minimal in energy due to the surface gradient energy term $F_{\rm SG}$. The texture can lower its energy by flipping the angle $\beta$ at the surface and the transition to the \textit{extended} flare out texture occurs\footnote[1]{There is ambiguity in the terminology in different publications: a) \textit{flare out} and \textit{flare in} is sometimes used instead of \textit{parted} and \textit{extended} flare out; b) \textit{flare out} sometimes refers to both \textit{parted} and \textit{extended} flare out texture \cite{solntsev_2007,salomaa_1990,korhonen_1990} }. Here $\beta$ now passes across $90^\circ$ in the vicinity of the surface boundary. Further increase of the applied flow extends the region with a near constant $\beta$ even more, until saturation.

\section{Nuclear Magnetic Resonance}

The details of the texture would be hidden from observation were it not for the fact that spin-orbit interaction with the magnetic moment of the $^3$He nucleus allow us to probe the texture non-invasively \cite{kopu_2000}. The Leggett equations
\begin{eqnarray}
	\frac{\partial\vec{\theta'}}{\partial t} & = &  - \gamma\vec{H}_{\rm tot} + \frac{\gamma^2}{\chi_{\rm B}}\vec{S} \\
	\frac{\partial \vec{S}}{\partial t} & = & \gamma\vec{S}\times\vec{H}_{\rm tot} - \frac{\chi_{\rm B}}{\gamma^2} \Omega_{\rm B}^2\hat{\vec{n}}(\hat{\vec{n}}\cdot\vec{\theta'})
	\label{eqn: leggett equation spin time dependence}
\end{eqnarray}
describe the response to small rotations $\vec{\theta'}$ of the spin density $\vec{S}$, where 
$\chi_{\rm B}(T)$ is the temperature dependent B-phase susceptibility, $\Omega_{\rm B}(T)$ the temperature dependent B-phase longitudinal resonance frequency and $\vec{H}_{\rm tot}$ the sum of the static field and a small transverse rf field $\propto e^{i\omega t}$. In the high-field limit ($H \gg2.5\,{\rm mT}$), the transverse resonance frequency reduces to
\begin{eqnarray}
	\omega \approx \sqrt{\omega_{\rm L}^2+\Omega_{\rm B}^2\sin^2\beta} \approx \omega_{\rm L} + \frac{\Omega_{\rm B}^2}{2\omega_{\rm L}}\sin^2\beta
	\label{eqn: larmor frequency shift}
\end{eqnarray}
with $\omega_{\rm L} = |\gamma | H$ the Larmor frequency. Here we primarily look at the frequency shift from the Larmor value. In \textit{continuous wave} (cw) NMR experiments, the NMR response is measured as a function of the longitudinal magnetic field $H$. The spectrum is then converted to the frequency domain using the relation
\begin{eqnarray}
	\Delta \nu = \frac{ (H_{\rm L} - H) H }{ H^2_{\rm L} } \nu_{\rm rf} {\rm ,}
\end{eqnarray}
where $\nu_{\rm rf}$ is the frequency of the transverse excitation field, $H$ is the measured field and $H_{\rm L}$ the Larmor field. The latter can be obtained from the NMR response in the normal phase. We define the \textit{reduced} frequency shift $\Delta\tilde{\nu}$ using (\ref{eqn: larmor frequency shift})  as
\begin{equation}
	\Delta\tilde{\nu}\equiv \frac{2\omega_{\rm L}}{\Omega^2_{\rm B}}(\omega-\omega_{\rm L})=\sin^2\beta {\rm .}
	\label{eq: reduced freq shift}
\end{equation}

The Kramers-Kr\"{o}nig relation tells that the \textit{absorption} and \textit{dispersion} signals are interrelated and the area below the measured absorption line shape $\chi(\nu)$ is a constant. Measured NMR line shapes can thus consistently be compared with \eg calculations in different domains: in the frequency domain $\nu$ or in the reduced frequency domain $\tilde{\nu}$. The normalization of the spectra is done by dividing with the integrated area under the line shape above the baseline:
\begin{eqnarray}
	\chi'( \nu ) = \frac{\chi(\nu)}{\int \chi(\nu) d\nu} {\rm ,}\\
	\tilde{\chi}'( \tilde{\nu} ) = \frac{\chi(\tilde{\nu})}{\int \chi(\tilde{\nu}) d\tilde{\nu}} {\rm .}
\end{eqnarray}
The normalized NMR line shape $\tilde{\chi}'$ in the reduced frequency domain $\Delta\tilde{\nu}$ is free of the NMR setup parameter $\nu_{\rm rf}$.
For examples of conversions between the magnetic field $H$, frequency shift $\Delta \nu$ and reduced frequency shift $\Delta \tilde{\nu}$, see the different horizontal axes of the NMR line shapes in \fig \ref{fig:textures and nmr line shapes}.

\section{Numerical Calculations}
Numerical calculations have been performed by finding the equilibrium texture, \ie the texture which corresponds to the minimum configuration of the free energy and by determining the corresponding distribution of NMR intensity \cite{kopu_2007}. The calculations use the following input: for geometrical parameters the cylinder radius $R$, rotation velocity $\Omega$ and the number of vortices $N$, which for convenience we characterize by the rotation velocity $\Omega_{\rm v}$, at which this number of vortices is the equilibrium amount; for the environmental parameters the reduced temperature $T/T_{\rm c}$ and pressure $p$ is used; for the NMR parameters the Larmor frequency $\nu_{\rm L}$ and the inhomogeneity of the applied magnetic field $\Delta H/H$; and for the physical parameter the field velocity parameter $\lambda_{\rm HV}$. Other physical parameters are calculated directly from theory using temperature $T$ and pressure $p$, or taken from experiment (see \fig 1...4 of \cite{thuneberg_2001}).

The outline of the calculation is as follows: from the geometrical parameters the vortex configuration is determined and the counterflow profile $v_{\rm cf}(r)$ is calculated. Using (\ref{eq: orientational free energy})...(\ref{eq: field velocity energy}) together with the environmental parameters, the textural energy is minimized using the truncated Newton algorithm \cite{tn_2010} which gives the solution as the $\hat{\vec{n}}$-vector expressed in $\alpha(r)$ and $\beta(r)$. The discretization step for $r$ is $15\mu$m. 

For the determination of the NMR absorption spectrum, the so-called \textit{local oscillator model} is used: the fluid as an assembly of uncoupled oscillators with frequencies determined by the local value of $\beta(r)$, see (\ref{eqn: larmor frequency shift}). The NMR line shape is then a sum of the individual contributions in the volume $V$ 
\begin{eqnarray}
	f(\omega) = \frac{1}{V} \int d^3\vec{r} \delta[\omega-\omega(\vec{r})] {\rm .}
	\label{eqn: local oscillator model distribution}
\end{eqnarray}
The local inhomogeneity of the magnetic field $\Delta H/H$ causes dephasing on the local oscillators due to the spread $\Delta H$ around the average field value $H$. The characteristic time $\tau_{\rm H}$ is of order
\begin{eqnarray}
	\tau_{\rm H} = \left ( \omega_{\rm rf} \frac{\Delta H}{H} \right ) ^{-1} = {\rm \Gamma}^{-1}_{\rm H} {\rm .}
\end{eqnarray}
By introducing the term $-{\rm \Gamma_H}\vec{S}_\perp$ 
on the right-hand side of (\ref{eqn: leggett equation spin time dependence}), where $\vec{S}_\perp$ is the transverse component of the magnetization, the delta function in (\ref{eqn: local oscillator model distribution}) becomes effectively a Lorentzian with a width $\Delta \omega = {\rm \Gamma_H}$. We determined the average field inhomogeneity for $\Delta \omega$ by measuring the NMR response in the normal phase. The superfluid texture is of the flare out type: the angle $\beta$ changes as a function of radius $r$, hence the measured NMR line shape of the flare-out texture depends on the field inhomogeneity at $r$. In combination with that the NMR pickup coil does not probe the fluid homogeneously, it makes the field inhomogeneity as measured in the normal phase only indicative. Therefore, the field inhomogeneity parameter is a fitting parameter when measured and calculated spectra are compared.

The first order texture transition from the \textit{parted} to the \textit{extended} flare out texture and back, shows hysteresis in the dependence on $\Omega$ and $T$, therefore dual solutions exist for the texture around the critical velocity $\Omega_{\rm c2}$. To cope with this degeneracy, the program accepts $\alpha(r)$ and $\beta(r)$ as initial starting values and then proceeds to find the closest local minimum of the textural free energy. The initial guess can be as crude as a straight line from $\beta(0)=0$ to $\beta(R)=63.4^\circ$ or $\beta(R)=180^\circ -63.4^\circ$, for the \textit{parted} and \textit{extended} flare out textures respectively.

\section{Experimental Techniques}
\label{section: experiments}

NMR measurements were performed with two different setups, which both included a long cylindrical container with radius $R=3\,$mm mounted on the \textit{nuclear stage} of a rotating cryostat. The cryostat rotates up to 3\,rad/s with an $\Omega$-dependent noise level which is $\Delta \Omega=0.006\,$rad/s (peak-to-peak) at 1\,rad/s.

\begin{figure}[!t]
	\begin{center}
		\includegraphics[width=0.3\textwidth]{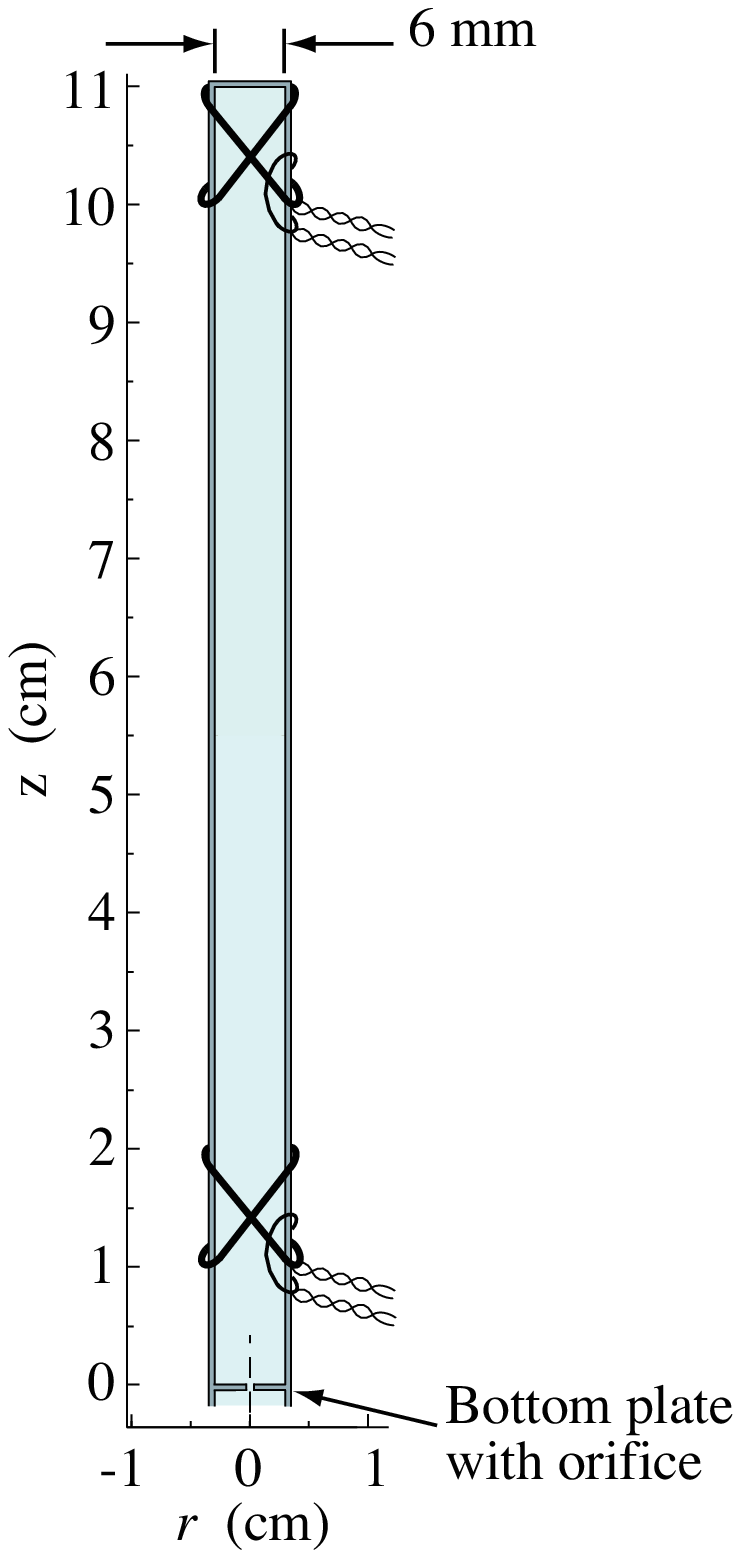}
		\includegraphics[width=0.6\textwidth]{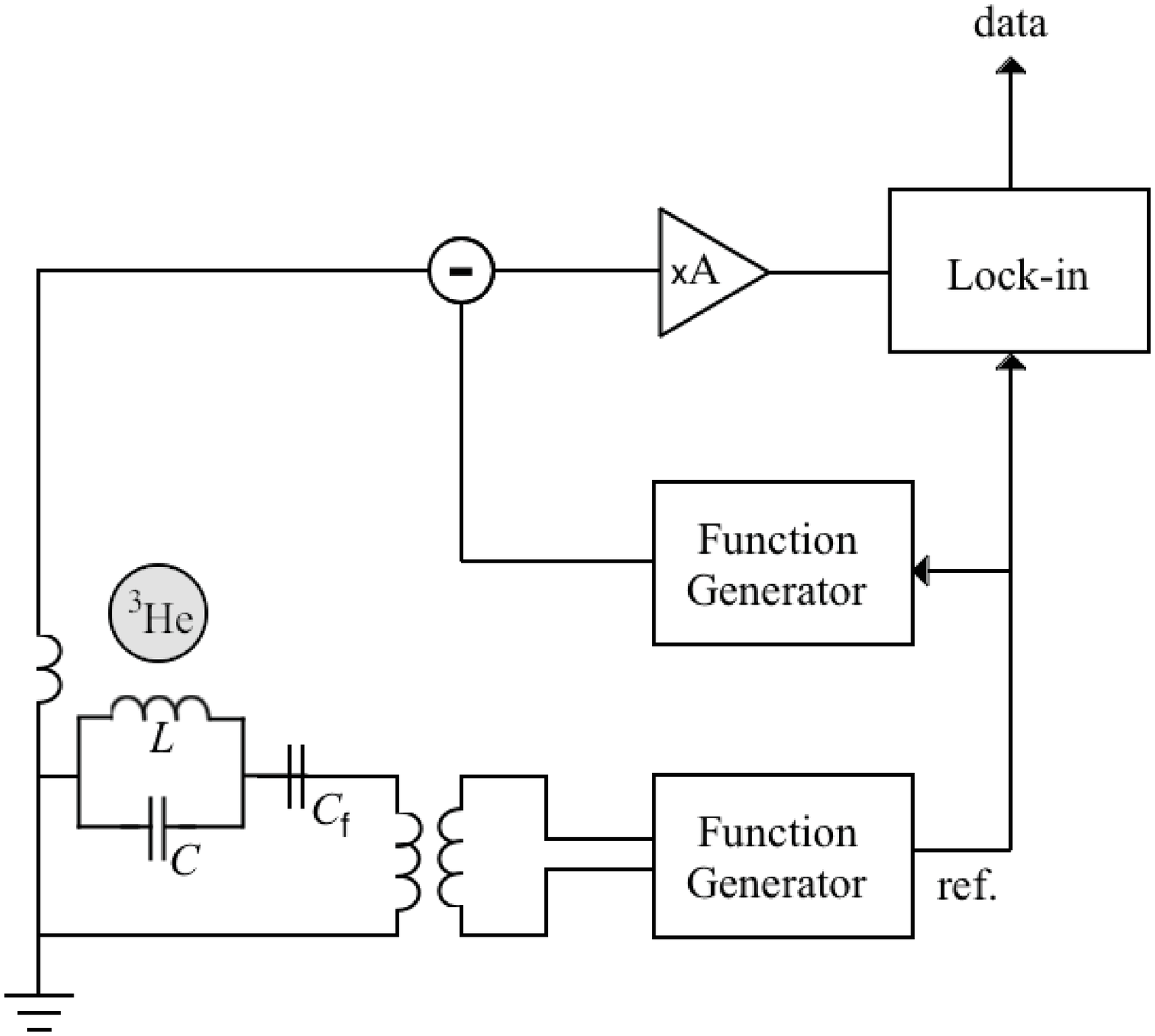}
	\caption{\textbf{Experimental setup in 2005}: the left figure shows the experimental quartz cylinder with a detector coil at both ends of the NMR sample volume. The bottom spectrometer was used for the linear NMR (full spectra) measurements, while the top spectrometer was used for non-linear NMR (magnons) measurements. The field inhomogeneity was determined by fitting the superfluid NMR line shape with calculations and was  consistently $\Delta H/H\approx8.8\cdot10^{-4}$ at all measured temperatures. 
The right figure shows schematically the LC-resonance circuit and the NMR Q-meter setup. The superfluid is excited with a high-Q LC-resonator; the magnetic pickup changes the inductance which is measured by the pickup loop. The signal is compensated by a second phase locked function generator for high resolution measurement after amplification using the Stanford Research Systems SR560 differential amplifier. A phase lock-in amplifier (Stanford Research Systems model SR844) is used for data acquisition.}
	\label{fig:experimental setup}
	\end{center}
\end{figure}

The \textit{2005 setup} (referring to the starting year) had above the sinter on top of the nuclear stage two quartz tuning fork oscillators (henceforth forks). The main sample volume is separated from the fork volume by a plate with an orifice ($\O=0.75\,$mm). The plate prevents vortices from entering the main volume. At each end of the main volume, two NMR pick up coils were mounted for \textit{continuous wave} (cw) NMR. See \fig \ref{fig:experimental setup} for the cell and the NMR resonance circuit.  For the flare out texture measurements in this paper, we use the bottom spectrometer. The order parameter texture was probed using a resonance circuit tuned to $\nu_{\rm rf}=965.0\,$kHz with a quality factor $Q=6050$ and a Larmor field of $H_{\rm L}=29.75\,$mT.

The \textit{2009 setup} had two forks in an isolated volume, which sits between the sinter volume and the main volume. Two orifices connect the three volumes. The orifice on the main volume side was bigger than the orifice on the sinter side, such that at low temperatures the quasi-particle flux from the main volume side dominates the quasiparticle density of the fork volume \cite{eltsov_2010_2}. The NMR coil at the bottom end was used for full spectra measurements. The resonance circuit had a pre-amp at 4\,K and a second amplifier at room temperature. The resonance frequency of $\nu_{\rm rf}=1.967\,$MHz was used, which corresponds to the Larmor field $H_{\rm L}=60.65\,$mT. The resonance circuit had a quality factor of $Q=3900$. The transverse rf field of the spectrometer coil was $H_{\rm rf}=3.4\,{\rm nT}$ and the coil had an inductance of $L=12.7\mu{\rm H}$. 
%
  
In both setups the temperature was measured using a fork, which was calibrated above $0.35\,T_{\rm c}$ against a Melting Curve Thermometer (MCT) mounted on the nuclear stage \cite{greywall_1986}. In the low temperature regime where the MCT loses thermal contact with the $^3$He NMR sample, the resonance width of the fork response was interpreted to be proportional to the quasi-particle density which depends exponentially on temperature \cite{vollhardt_1990}. The heatleak was determined to be $\dot{Q}=20$pW (at $\Omega=1\,{\rm rad/s}$) for the \textit{2009 setup}, while for the \textit{2005 setup} this was estimated to be of one order of magnitude larger.

The superfluid was continuously probed by sweeping slowly the magnetic field in the NMR magnet and measuring the absorption $\chi(H)$ and dispersion in the NMR pickup coils at 4 samples/s with a lock-in amplifier. The typical sweep rate of the main NMR field was $\dot{H}=9\div13\mu{\rm T/s}$.

The magnetic field of the NMR is produced by a combination of solenoid coils which by design have an axial and radial inhomogeneity of $\Delta H_{\rm z}/H_{\rm z} = 7.3\cdot 10^{-5}$ and $\Delta H_{\rm r}/H_{\rm r} = 2.0\cdot 10^{-5}$, respectively. Niobium shields protect against external magnetic influences. In practice the field inhomogeneity is increased by an order of magnitude by the presence of the superconducting NMR pickup coils. The field inhomogeneity was measured a number of times (after different cool downs to liquid helium temperatures) in the Fermi-liquid state (normal phase) with low enough excitation not to saturate the signal.
The NMR response was distorted and far from an ideal Lorentzian. A poor fit gave an inhomogeneity of $1.7\cdot10^{-3}$.  Measurements of the field inhomogeneity were taken with different magnetic fields $H_{\rm demag}=0.1\div 1.1{\rm T}$ in the superconducting magnet used for adiabatic demagnetization cooling (or nuclear cooling), to verify that the superconducting Nb shields around the NMR setup are protecting the NMR measurement from external magnetic influences. The discussion on the field inhomogeneity is continued in \sect \ref{sec: Comparison of Measurements with Calculated Textures} where we use the inhomogeneity as a fitting parameter.

The NMR line shapes were processed by substracting the baseline and correcting the phase using both the measured \textit{absorption} and \textit{dispersion} signals of the lock-in amplifier. The magnetic field sweep was linearly corrected for the inductive delay of the magnet at sweep rate $\dot{H}$; this correction made up and down sweeps to overlap.

Since we want to measure the NMR line shape of the vortex free state as well as those with different number of rectilinear vortices $N$, the sample state is prepared above the onset temperature $T_{\rm on}$ of  turbulence \cite{solntsev_2007,de_graaf_2008}. A rotating vortex free state is created by increasing the rotation velocity $\Omega$ starting from zero with $\dot{\Omega}=0.03\,{\rm rad/s^2}$, at a temperature $T=0.7\div0.75\,T_{\rm c}$. This temperature is above the onset temperature $T_{\rm on}$ to turbulence and below the AB-phase transition temperature $T_{AB}$. To prepare the superfluid with a predefined vortex cluster, the following recipe was used: at these high temperatures the superfluid is rotated above the critical rotation velocity $\Omega_{\rm cr}$ (in our cell $\Omega_{\rm cr} = 2.5\div3.5$rad/s) where vortices are created when the cf velocity reaches the bulk liquid critical value at the largest imperfections on the cylindrical surface. At these high velocities the cf-peak drops significantly indicating that a large number of vortices are created. The rotation is then slowed down to the desired rotation velocity $\Omega_{\rm v}$. After annihilation of extra vortices and reaching the equilibrium vortex state, the cylinder contains $N_{\rm v}$ rectilinear vortices. The rotation velocity is then increased to a desired value $\Omega < \Omega_{\rm cr}$ making sure that no new vortices are created. The cell is now rotating at a rotation velocity $\Omega$ with a known number of vortices $N$, which is smaller than the equilibrium number of vortices at this rotation velocity \cite{eltsov_2010}.

The measurements in the \textit{2005 setup} were performed by recording the NMR line shape during a slow cool down ($\dot{T} \approx 50\cdot10^{-6}\,T_{\rm c}$/s) to low temperatures. The cool down was frequently interrupted for temperature stabilization/verification. The rotation velocities used were $\Omega=0.7, 0.8, 0.9, 1.05\,$rad/s, with clusters of vortices corresponding to $\Omega_{\rm v}=0, 0.1, 0.25$rad/s. These low numbers of vortices have a significant effect on the texture \cite{kopu_2000}. 
After reaching low temperatures, the temperature was swept back to the starting temperature to compare the initial and final NMR line shapes to make sure that no new vortices had been created in this temperature cycle. Note that this cycle takes place to a large extent in the regime where superfluid $^3$He-B can have turbulent dynamics: below $T\approx0.6\,T_{\rm c}$ the superfluid Reynolds number $Re_{\alpha}$ increases rapidly from unity to $Re_{\alpha}=10^3$ at $T=0.2\,T_{\rm c}$ \cite{bevan_1997}. However, the experiments presented in this paper are not affected by turbulence, due to the fact that there are no vortices connected to the cylindrical boundary \cite{eltsov_2010}.

In the measurements taken in the \textit{2009 setup}, the cell was prepared in the vortex free state by increasing the rotation velocity at $T\approx 0.7\,T_{\rm c}$ without the creation of vortices. The cell was then cooled down at constant rotation $\Omega_{\rm i}$ to a preselected temperature $T$ where the NMR line shapes were measured in the vortex free state at different rotation velocities $\Omega$ in the range $0.35\div2\,$rad/s. Between the measurements at the varying rotation velocities $\Omega$, the rotation was brought back to the initial rotation velocity $\Omega_{\rm i}$ for verification that no new vortices had been created by comparing the initial and current NMR line shapes. Temperature control was stable within $\Delta T = 0.005\,T_c$, as measured by the forks. 

\begin{figure}[!t]
	\begin{center}
		\includegraphics[width=0.9\textwidth]{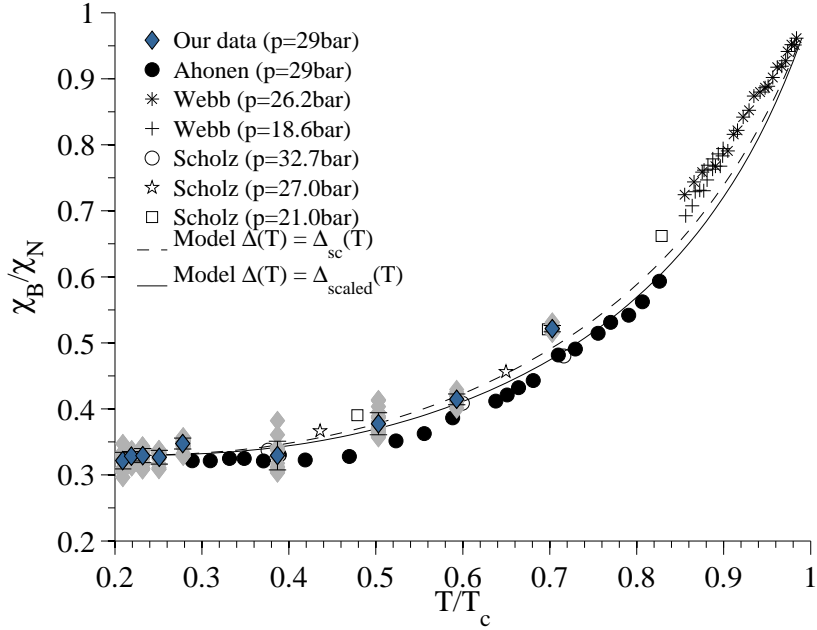}
		\caption{\textbf{Magnetic susceptibility of $^3$He-B}: The reduced magnetic susceptibility $\chi_{\rm B}/\chi_{\rm N}$ as a function of reduced temperature $T/T_{\rm c}$. The results are obtained by taking the ratio between the integrated (baseline, phase and amplitude corrected) NMR line shape in the vortex free superfluid state (Landau state) and the Fermi-liquid state (normal state). Measurements are shown taken at different $\Omega$ (gray markers) in the range $\Omega=0.15\div2.01$rad/s. The scatter results from the baseline drift during the measurement. There was no cross-correlation between the scatter in $\chi_{\rm B}/\chi_{\rm N}$ and $\Omega$, \ie random scatter around the average. Averages (blue markers) are taken on isothermals with the error-bars indicating one standard deviation from the average value. Our measurements are compared with measurements by Webb \etal \cite{bozler_1992} and by Scholz \cite{scholz_1981} published in reference \cite{inseob_hahn_1998}. The two curves represent the theoretical model (\ref{eq: reduced susceptibility}) with different values for the energy gap $\Delta$ which enters through the Yosida function. The solid line shows the model with $\Delta_{\rm sc}(T)$ and the dashed line with a scaled energy gap $\Delta_{\rm scaled}(T)$ (see (\ref{eq: gap sc}) and (\ref{eq: gap scaled}) for definitions). }
	\label{fig:magnetic susceptibility}
	\end{center}
\end{figure}

\section{Dynamic Susceptibility Measurement}

We use the area of our spectra as a measure for the \textit{static} susceptibility by comparing the absorption $\chi(\omega)$ in the superfluid state with the absorption $\chi_{\rm N}$ in the normal state above $T_{\rm c}$ to assure linear NMR was measured. The signal amplitude relates to the susceptibility $\chi_{\rm B}(T)$ as
\begin{eqnarray}
	\frac{\chi(\omega)}{\chi_{\rm N}} = \frac{\omega_o \pi}{2} \frac{\chi_{\rm B}(T)}{\chi_{\rm N}} \frac{V_{\rm s}(\omega)}{\int V_{\rm s}(\omega')d\omega'} {\rm ,}
\end{eqnarray}
where $\chi_{\rm B}(T)/\chi_{\rm N}$ is defined as the ratio of the total integrated NMR absorptions in the superfluid phase and the normal phase which can be experimentally determined
\begin{eqnarray}
	\frac{\chi_{\rm B}(T)}{\chi_{\rm N}}= \frac{(\int V_{\rm s}(\omega')d\omega')_T}{(\int V_{\rm s}(\omega')d\omega')_{T_{\rm c}}} {\rm ,}
\end{eqnarray}
The results of $\chi_{B}(T)$ measurements of the vortex free state at different rotation velocities $\Omega$ as a function of temperature $T$ are plotted in \fig \ref{fig:magnetic susceptibility}. The values of $\chi_{\rm B}$ at different $\Omega$ have a larger scatter, but there was no correlation with the applied velocity $\Omega$. During the measurements the baseline was drifting,  which affects the normalization and is seen as the main cause of the scatter. The theoretical expression for the susceptibility according to the weak coupling theory is \cite{dobbs_2000}
\begin{eqnarray}
	\frac{\chi_{\rm B}}{\chi_{\rm N}} = (1 + F^{\rm a}_0)\frac{\frac{2}{3} + \frac{1}{3}Y(T)}{1+F^{\rm a}_0(\frac{2}{3}+\frac{1}{3}Y(T))}
	{}_{\;\longrightarrow}^{T\rightarrow 0}
	\frac{2(1+F^{\rm a}_0)}{3+2F^{\rm a}_0} {\rm .}
	\label{eq: reduced susceptibility}
\end{eqnarray}
At a pressure of 29bar, $F^{\rm a}_0(29)=-0.75$, which gives $\chi_{\rm B}/\chi_{\rm N}=0.33$.

In \fig \ref{fig:magnetic susceptibility}, $\chi_{\rm B}/\chi_{\rm N}$ is plotted for the energy gaps $\Delta_{\rm sc}(T)$ and $\Delta_{\rm scaled}(T)$. The gap $\Delta_{\rm sc}$ is the weak-coupling gap $\Delta_{\rm wc}(T)$ with corrections using the \textit{strong-coupling} approximation. The scaled gap $\Delta_{\rm scaled}(T)$ is the \textit{strong-coupling} corrected gap $\Delta_{\rm sc}$ linearly scaled such that it coincides in the low temperature limit with the experimental observed gap $\Delta_{\rm exp}$ as measured by Todoshchenko \etal \cite{todoshchenko_2002}. To summarize:
\begin{eqnarray}
	\Delta_{\rm wc}(T)    &=& {\rm gap\;weak-coupling} {\rm ,}
	\label{eq: gap wc} \\
	\Delta_{\rm sc}(T)    &=& \Delta_{\rm wc}(T) |_{\rm strong-coupling\;corrected} {\rm ,}
	\label{eq: gap sc} \\
	\Delta_{\rm scaled}(T)&=& \Delta_{\rm sc}(T) \frac{\Delta_{\rm exp}(0)}{\Delta_{\rm sc}(0)} {\rm .}
	\label{eq: gap scaled}
\end{eqnarray}
The low temperature limiting values of the gap at $p=29\,$bar are $\Delta_{\rm wc}(0)=1.79\,k_{\rm B}T_{\rm c}$, $\Delta_{\rm sc}(0)=1.87\,k_{\rm B}T_{\rm c}$ \cite{thuneberg_2001} and $\Delta_{\rm exp}(0)=1.97\,k_{\rm B}T_{\rm c}$.

The plot includes high temperature data from Webb \etal \cite{bozler_1992} and Scholz \cite{scholz_1981} at a variety of pressures. Both from their data and ours it is seen that the pressure dependence of $\chi_{\rm B}/\chi_{\rm N}$, which comes about via $F^{\rm a}_0$ and the strong-coupling corrections, is not very strong above $p=20\,bar$. Also included is data from Ahonen \etal \cite{ahonen_1976} which is taken at the same pressure ($29\,$bar) as our measurement. The determination of the \textit{static} susceptibility of superfluid $^3$He using the imaginary component of the \textit{dynamic} susceptibility measurements (NMR) \textit{integrated over all frequencies}, has earlier been in conflict with "true" \textit{static} SQUID-based susceptibility measurements. Hahn \etal \cite{inseob_hahn_1998} suggest the difference with the NMR measurements, where the integrated area is used to determine the susceptibility, the integration range is limited to frequencies around Larmor $\nu_{\rm L}$ and the SQUID measurements include an extra contribution caused by, \eg the electrons. For now, our measurements (at $T<0.7\,T_{\rm c}$) of the \textit{static} susceptibility using NMR is consistent with the NMR data of Ahonen \etal and Scholz.

\section{Textural Transitions and CF-Peak Behavior}
\label{sec: Analysis and Results}

In measurements at fixed temperature, we have observed textural transitions by changing the rotation velocity $\Omega$ by a step (rapid acceleration at $\dot{\Omega}=0.030\,$rad/s). The critical velocity $\Omega_{\rm c1}$ of the second-order transition from the \textit{simple} to the \textit{parted} flare out texture is difficult to determine exactly since the transition is continuous.

\begin{figure}[tp]
	\begin{center}
		\includegraphics[width=0.7\textwidth]{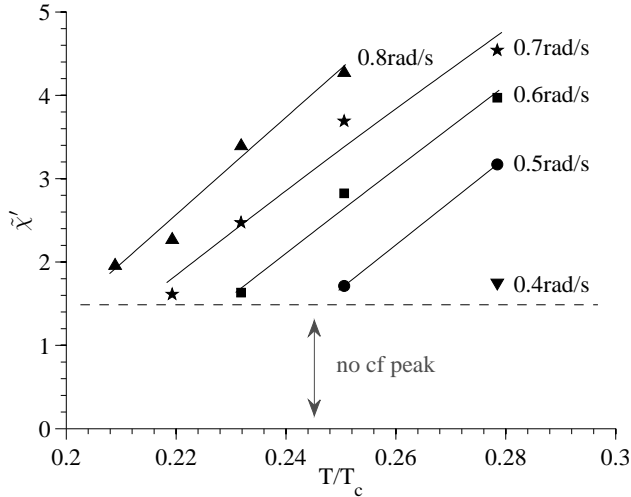}
		\caption{ \textbf{Counterflow peak behavior near $\Omega_{\rm c1}$}: cf-peak height $\tilde{\chi}_{\rm cfp}'$ plotted as a function of reduced temperature given in terms of the rotation velocity $\Omega$ in the vortex-free state. In the plot the cf-peak starts to appear at $\tilde{\chi}_{\rm cfp}'\approx1.5$, where the texture transforms from \textit{simple} to the \textit{parted} flare out texture. Lines connect data points with the same applied flow. With decreasing temperature, the measured \cf-peak height decreases rapidly.}
		\label{fig:absorption at low temperature}
	\end{center}
\end{figure}

\fig \ref{fig:absorption at low temperature} shows the cf-peak height $\tilde{\chi}_{\rm cfp}'$ (plotted in the reduced frequency domain) as a function of temperature for the \textit{parted} flare out texture with no vortices present at several rotation velocities $\Omega$. At $\tilde{\chi}'\approx 1.5$ is the transition line where the cf-peak vanishes. What is left is the absorption of the \textit{simple} flare out. The transition $\Omega_{\rm c1}$ does not seem to have any hysteresis within the resolution of the step $\Delta\Omega=\pm 0.050\,$rad/s we used.

In contrast, the second texture transition at $\Omega_{\rm c2}$ between the \textit{parted} and \textit{extended} flare out textures has a large hysteresis in the $\Omega_{\rm}$-dependence and is of first-order. This transition happens instantaneously within our sampling rate of four samples/s. The hysteresis increases with decreasing temperature: around $T=0.5\,T_{\rm c}$ the hysteresis is within $\Delta\Omega=0.3\,$rad/s, while at $T=0.25\,T_{\rm c}$ the hysteresis can exceed $\Delta\Omega=0.8\,$rad/s. Typically, the texture change seems to happen during the step change in rotation velocity and rarely occurred during stable rotation.

\begin{figure}[!t]
	\begin{center}
		\includegraphics[width=0.7\textwidth]{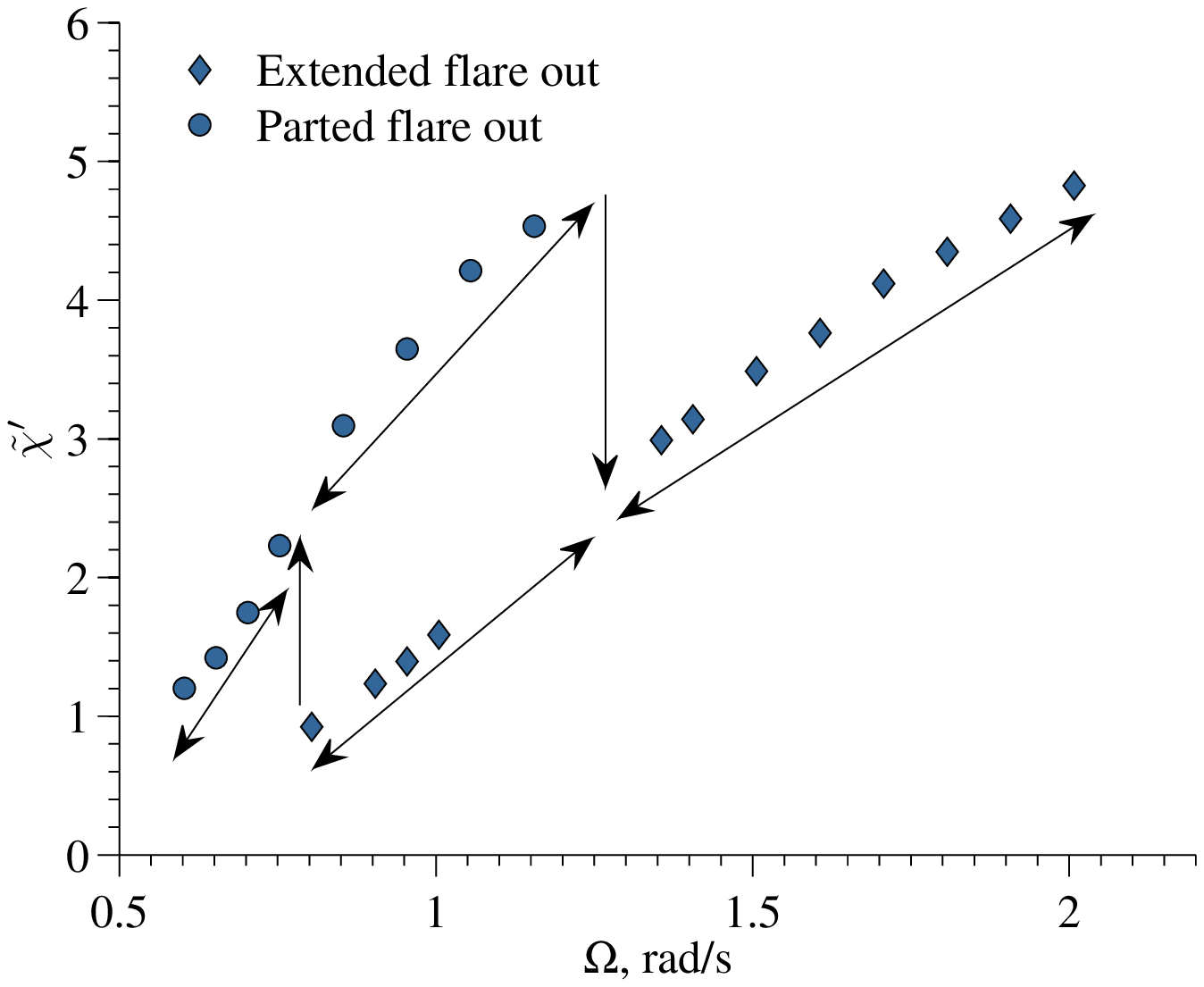}
		\caption{\textbf{Velocity hysteresis of the critical texture transition $\Omega_{\rm c2}$ between the \textit{parted} and \textit{extended} flare out texture:} the plot shows the cf-peak at $T=0.25\,T_{\rm c}$ with a vortex cluster of $\Omega_{\rm v} = 0.12\,$rad/s during a velocity sweep down starting at 1\,rad/s and up. In this cycle, the NMR spectra were recorded and the cf-peak was extracted after each spectrum was normalized in the reduced frequency domain. A transition occurred from the \textit{extended} to the \textit{parted} flare out texture when the velocity $\Omega$ was lowered to $\Omega=0.8\,$rad/s. When the velocity was lowered further, the texture did not change. After step wise  increasing $\Omega$, the texture stayed in the \textit{parted} flare out texture until $\Omega$ was increased to $1.35\,$rad/s where the texture flipped back to the \textit{extended} texture.}
		\label{fig:omega sweep around omega c2}
	\end{center}
\end{figure}
\begin{figure}[!h]
	\begin{center}
		\includegraphics[width=0.7\textwidth]{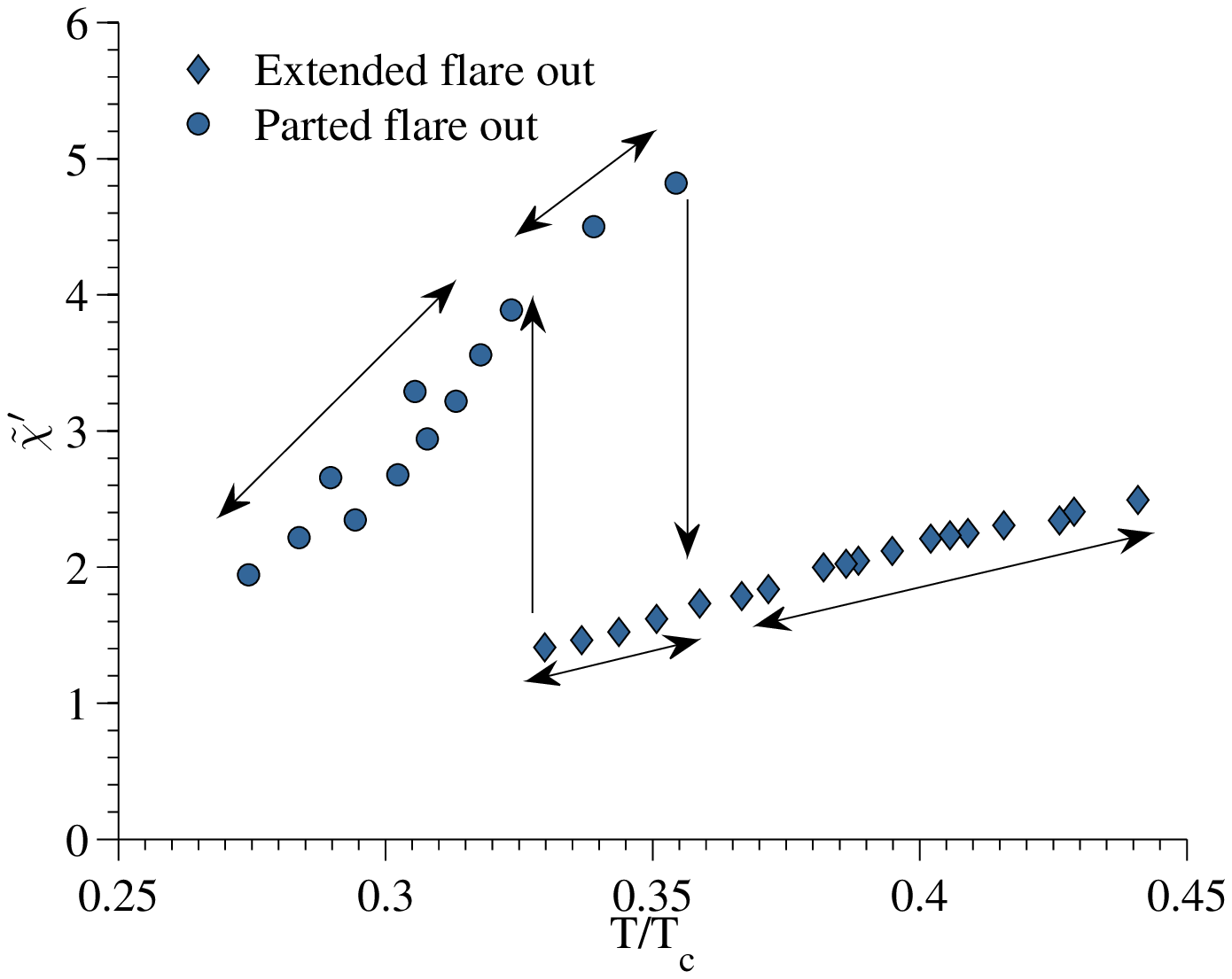}
		\caption{\textbf{Thermal hysteresis of the critical texture transition $\Omega_{\rm c2}$ between the \textit{parted} and \textit{extended} flare out textures}: the plot shows the cf-peak at $\Omega=0.8\,$rad/s with a vortex cluster $\Omega_{_v}=0.1\,$rad/s, during a temperature sweep from high to low temperatures and back. In this cycle, the NMR line shape was recorded and the cf-peak was extracted after each spectrum was normalized in the reduced frequency domain. At high temperatures the order parameter was in the \textit{extended} flare out texture and slowly cooled down. When the temperature cooled to $T=0.33\,T_{\rm c}$, the texture changed to the \textit{parted} flare out texture. After reaching the low temperature point of $T=0.27\,T_{\rm c}$, the temperature sweep direction was reversed. After reaching a temperature of $T=0.354\,T_{\rm c}$ as measured by the fork, the \textit{parted} flare out texture flipped back to the \textit{extended} flare out texture. Note that the relative noise of the cf-peak measurement increases in this figure with decreasing temperature due to the normalization of the spectra when the susceptibility decreases with decreasing temperature. }
		\label{fig:temp sweep around omega c2}
	\end{center}
\end{figure}

\fig \ref{fig:omega sweep around omega c2} shows the hysteresis of $\Omega_{\rm c2}$ in the velocity  dependence at $0.25\,T_{\rm c}$ with a vortex cluster of $\Omega_{\rm v}=0.12\,$rad/s. A transition from the \textit{extended} to the \textit{parted} flare out texture occurs around $\Omega\approx 0.75\,$rad/s when the velocity is step wise decreased. If the velocity is increased the texture change is reversed to the \textit{extended} flare out texture at $\Omega=1.35\,$rad/s.

In the measurements at constant $\Omega$ the temperature was swept. The texture transition at $\Omega_{\rm c2}$ between the \textit{parted} and \textit{extended} flare out textures is also observed to be hysteretic with a transition regime of $\Delta T= 0.025T_{\rm c}$ width. \fig \ref{fig:temp sweep around omega c2} shows the cf-peak behavior in a temperature cycle from high to low temperatures and back. In the low temperature regime, the temperature at which the transition from the \textit{extended} to \textit{parted} flare out texture occurs on a cool down is lower than the temperature of the transition from the \textit{parted} to the \textit{extended} flare out texture during a warm up. The large increase in absorption of the cf-peak from the \textit{extended} to \textit{parted} flare out texture during cooling is due to the $\beta$ angle not crossing the $90^\circ$ angle in the \textit{parted} flare out texture. The smaller $\beta$ angle enhances the absorption of the cf-peak since the absolute value of the free energy term $|F_{\rm HV}|$ is now smaller, and less energy is required to orient $\vec{l}$ along the direction of the flow. In effect $\vec{l}$ orientates along the flow over a larger region of space. Or in different words: the cf peak increases in the texture change from the \textit{extended} to \textit{parted} flare out texture, since the spectrum width becomes smaller. The argument is reversed for the change in cf-peak absorption in the texture transition from the \textit{parted} to the \textit{extended} flare out texture: the angle $\beta$ now crosses $90^\circ$ and is on average steeper over the same space and the applied flow affects the tilt of the $\beta$ angle less.

\begin{figure}[tp]
	\begin{center}
		\includegraphics[width=0.9\textwidth]{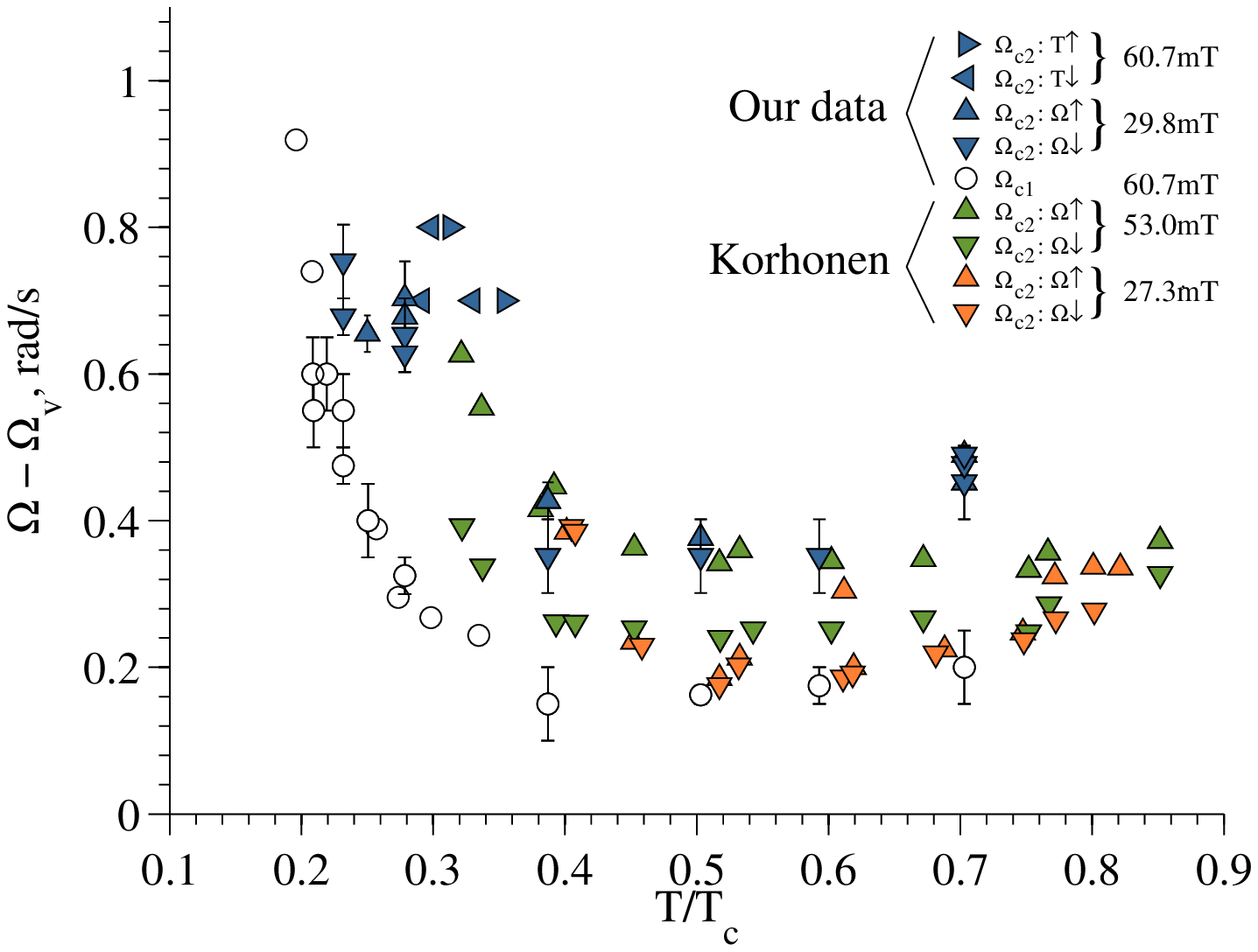}
		\caption{\textbf{Critical textural transitions}: figure shows the critical velocity $\Omega_{\rm c1}$ (between \textit{simple} and \textit{parted} flare out textures) and $\Omega_{\rm c2}$ (between \textit{parted} and \textit{extended} flare out textures) given as the cf velocity $\Omega_{\rm cf} = \Omega - \Omega_{\rm v}$ versus reduced temperature. The first critical velocity $\Omega_{\rm c1}$ is constant down to $T=0.4T_{\rm c}$ and then increases rapidly. The error bars indicate the magnitude of the velocity step $\Delta\Omega$ when the transition was measured. The second critical velocity $\Omega_{\rm c2}$ was measured in \textit{setup 2005} at constant rotation velocity by sweeping the temperature ($T\uparrow$: increase in temperature), and in the \textit{setup 2009} at constant temperature by varying rotation velocity ($\Omega\uparrow$: rotation is increased by a step $\Delta\Omega$). Error bars indicate the velocity step where the transition occurred. The pressure of the $^3$He liquid was $p=29\,$bar. Our results are compared with data from Korhonen \etal (1990) at pressure $p=10.3\,$bar \cite{korhonen_1990}. Larmor field values are given in the legend.}
		\label{fig:critical textural transitions}
	\end{center}
\end{figure}

In \fig \ref{fig:critical textural transitions} the textural transition $\Omega_{\rm c1}$ between the \textit{simple} and \textit{parted} flare out textures, as well as the textural transition $\Omega_{\rm c2}$ between the \textit{parted} and \textit{extended} flare out textures are plotted as a function of reduced temperature. At high temperatures down to $T\approx0.4\,T_{\rm c}$, the transition velocity $\Omega_{\rm c1}$ is approximately constant. At temperatures below $T\approx0.4\,T_{\rm c}$, $\Omega_{\rm c1}$ increases rapidly. The hysteresis in the transition velocity $\Omega_{\rm c2}$ is approximately constant ($\Delta\Omega=0.1\,$rad/s) down to $\approx0.4\,T_{\rm c}$ where it starts to increase rapidly with decreasing temperature. Below $0.25\,T_{\rm c}$, the hysteresis in applied counterflow is so big that once the system is in the \textit{parted} flare out texture a transition back to the \textit{extended} flare out texture was not observed below a critical velocity $\Omega_{\rm cr}\approx 2.1\,$rad/s at which vortices nucleated due to imperfections of the cylindrical surface or multiplication of trapped vortices in the corner of the cylindrical container. Our result is compared with data obtained by Korhonen \etal \cite{korhonen_1990} at similar magnetic fields $H$, but at differing liquid pressures: our experiment was at a pressure of $29\,$bar, while Korhonen's experiment was at $10.3\,$bar. Taking this pressure difference into account, the agreement is rather good.
 
\begin{figure}[tp]
	\begin{center}
		\begin{picture}(350,225)
			\put(0,0) { \includegraphics[width=0.8\textwidth]{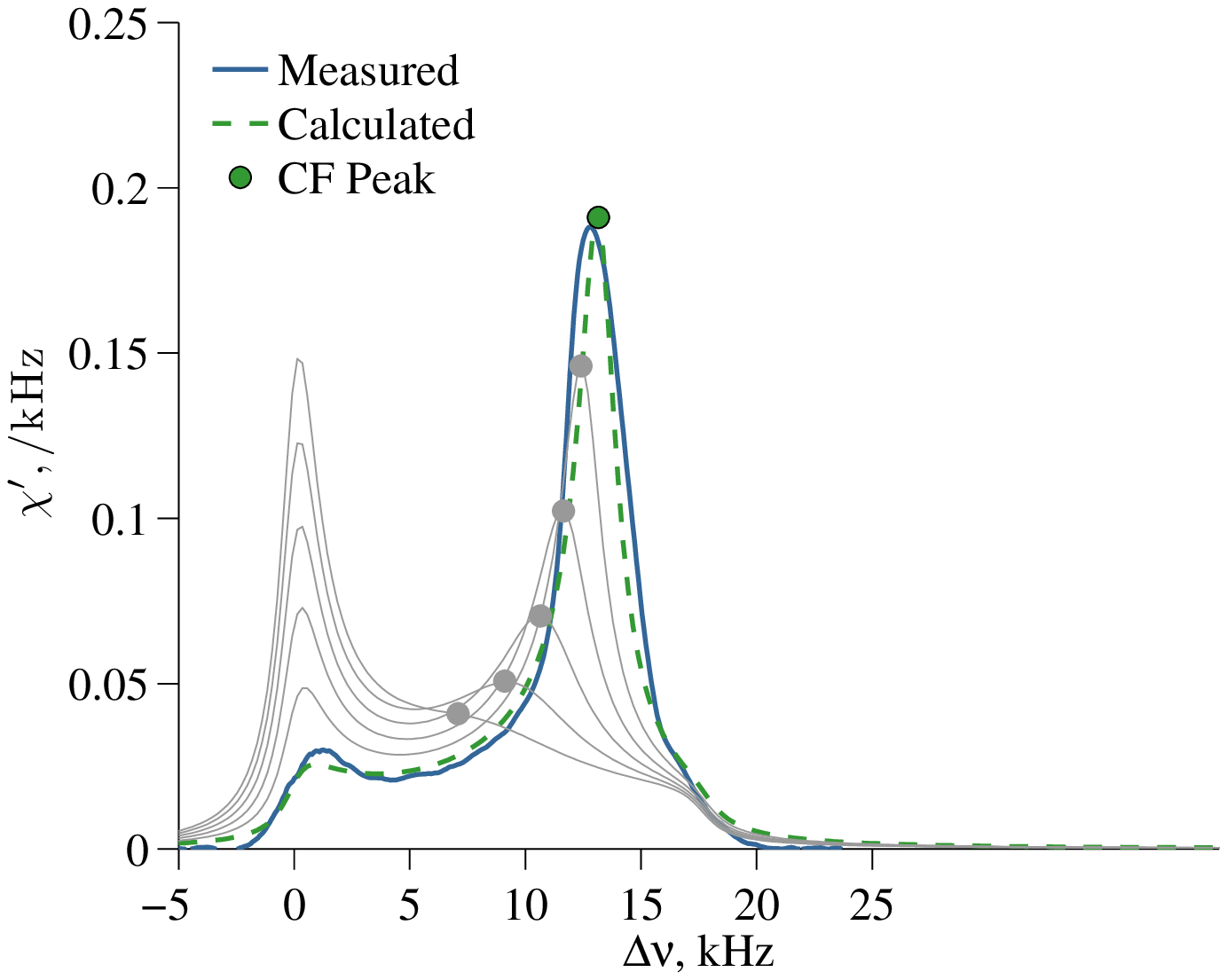} }
			\put(200,10){ \includegraphics[width=0.4\textwidth]{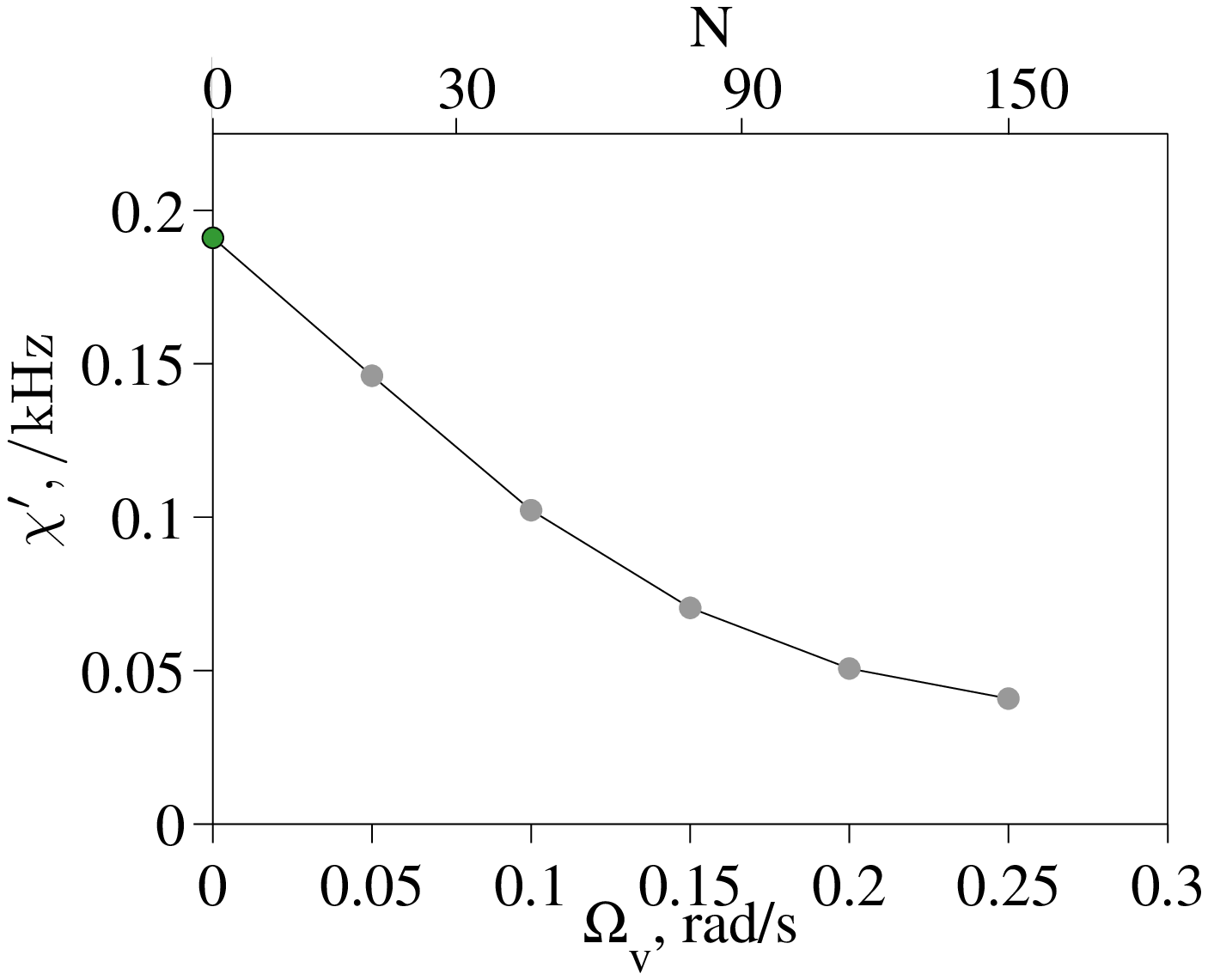} }
			\put(207,125){ \includegraphics[width=0.38\textwidth]{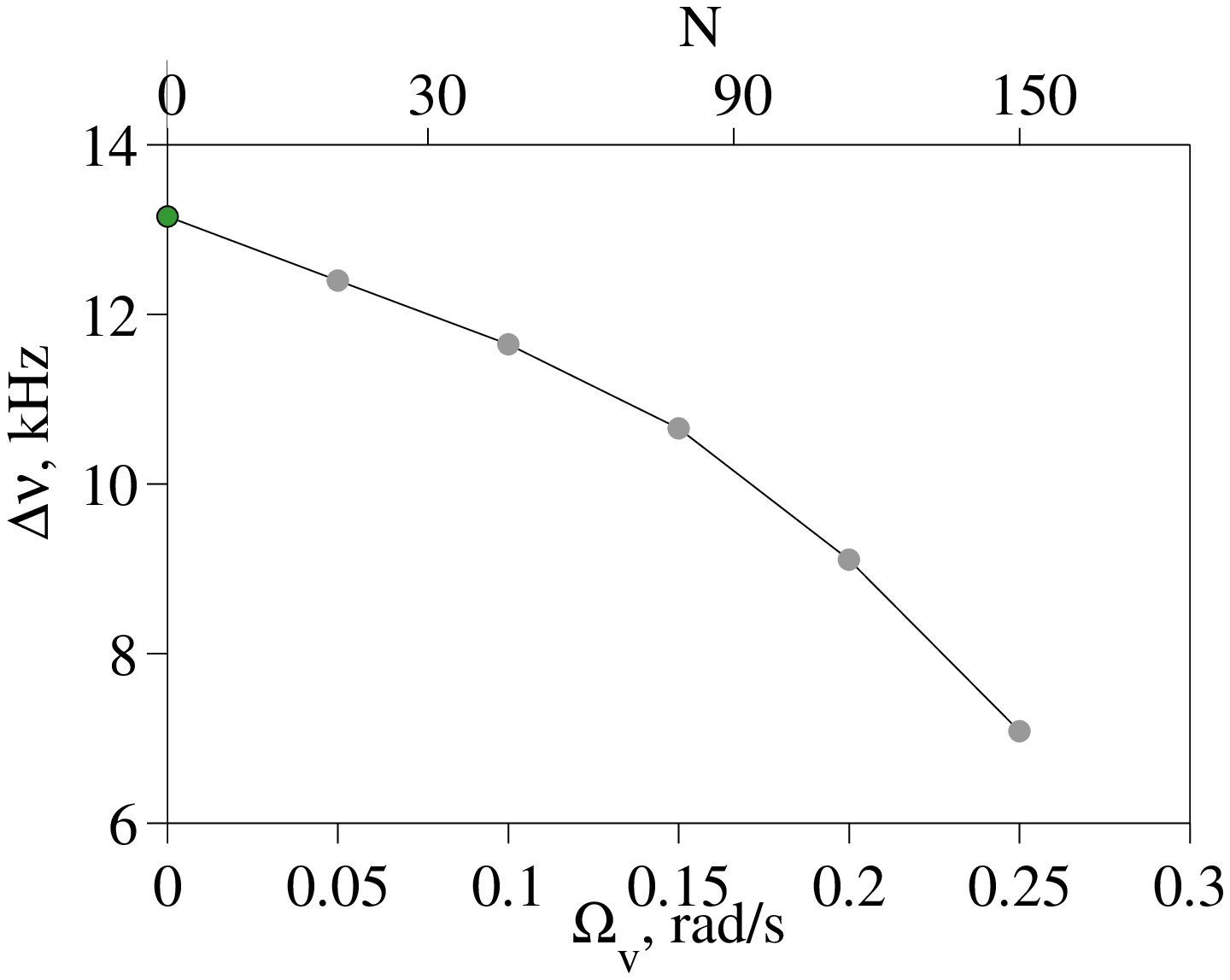} }
		\end{picture}
		\caption{\textbf{Spectra as a function of equilibrium vortex cluster $\Omega_v$}: The left figure shows the experimental (blue) and the optimal calculated (dashed green) line shape of vortex free superfluid $^3$He at $0.232\,T_{\rm c}$ rotating at $\Omega=0.9\,$rad/s. The experimental line shape is an average of 3 measured spectra. The NMR resonance frequency was at $\nu_{\rm rf}=1.97\,$MHz. The optimal value for the field velocity parameter was $\lambda_{\rm HV}=0.94\,$kg/m$^3$T$^2$. Gray lines show the calculated line shapes with an increasing amount of vortices; we increased the equilibrium vortex cluster $\Omega_{\rm v}$ with steps of $\Delta\Omega_{\rm v}=0.05\,$rad/s. The top right figure shows the frequency shift of the cf-peak as a function of the number of vortices in the central cluster, as expressed in terms of $\Omega_{\rm v}$ and $N$. The bottom right figure show the extracted cf-peak height of the calculated NMR line shapes as a function the $\Omega_{\rm v}$ and $N$.}
		\label{fig:spectra variable omega_v}
	\end{center}
\end{figure}

Generally, the height of the cf peak is not an intrinsic property of $^3$He-B, since the \cf-peak height depends in a nonlinear fashion on the field homogeneity and experimental geometry. In our measurements, qualitatively speaking, the \cf-peak height increases almost linearly in the \textit{parted} flare out texture with increasing rotation velocity starting from $\Omega_{\rm c1}$. At the texture transition from the \textit{parted} to the \textit{extended} flare out texture, the \cf-peak height drops significantly when compared to the \cf-peak height at the same rotation velocity in the \textit{parted} flare out texture. The increase of the cf-peak in the \textit{extended} flare out texture continues as a function of $\Omega$ until its height slowly saturates, \ie a maximum in absorption has been reached where all the area of the NMR line shape has moved "under" the \cf-peak. As for the frequency shift $\Delta\nu$ of the \cf-peak: in the \textit{parted} flare out texture, the positive frequency shift from the Larmor frequency increases with increasing cf velocity. The frequency shift continues to increase in the \textit{extended} flare out texture until it quickly saturates at $\Delta\tilde{\nu}=\sin^2\beta=\frac{4}{5}$. Examples of the absorption and frequency shift measurements of the \cf-peak have been published earlier and we refer to \fig 1 in \cite{eltsov_2010} and \fig 3 in \cite{eltsov_2010_2}.

Kopu \etal \cite{kopu_2000} performed an extensive study on the \cf-peak behavior and its dependence on the vortex number at high temperatures ($T>0.75\,T_{\rm c}$). Since the \cf-peak height depends strongly on the exponentially decreasing density anisotropy in the low temperature limit $T\rightarrow 0$, in combination with the increasing textural transition velocity $\Omega_{\rm c1}$, the usefulness of this NMR  measurement technique can be put into question. However, NMR measurements are still a useful technique down to $0.2\,T_{\rm c}$, when enough flow is applied ($\Omega_{\rm cf} > \Omega_{\rm c1} =0.9\,$rad/s). As an illustration, \fig \ref{fig:spectra variable omega_v} shows a NMR line shape measurement of the vortex free state of $^3$He-B at $0.23\,T_{\rm c}$ rotating at $\Omega=0.9{\rm rad/s}$: the cf-peak is significant. Calculations of spectra with varying sizes of vortex clusters are compared to this measurement in the plot and demonstrate that the cf-peak is still a sensitive measure of the number of vortices.

\section{Comparison of Measurements with Calculated Textures}
\label{sec: Comparison of Measurements with Calculated Textures}
The measured NMR lines shapes converted in the frequency domain were fitted to numerical calculations  using the same geometrical, environmental and NMR parameters as applicable for the experimental setup (see \sect \ref{section: experiments}). Measurements carried out during a continuous temperature sweep were accepted only if the rate of change of the temperature was within the absolute value $|\dot{T}| < 5 \cdot 10^{-5}\,T_{\rm c}/{\rm s}$ (includes stable temperature measurements). The measured and calculated spectra were plotted in the same figure on top of each other as seen in \fig \ref{fig:textures and nmr line shapes}. We considered the fit to be good when (a) the NMR spectrum edge at high frequencies, which is determined by the B-phase longitudinal resonance frequency $\Omega_{\rm B}(T,p)$, overlap with the edge in the measured spectrum; (b) the \cf-peaks are nearly identical in height and frequency shift; (c) the difference in the normalized integrated area $\Delta A$ (see (\ref{eq: area between spectra})) between the spectra is smaller than 0.1 (dimensionless in the reduced frequency domain). 

The fitting parameters are the temperature $T$, the field inhomogeneity $\Delta H/H$, and the field velocity parameter $\lambda_{\rm HV}$. The initial value for the temperature was taken from the fork reading, the field inhomogeneity from the NMR response in the Fermi-liquid state and the field velocity parameter $\lambda_{\rm HV}$ with $\Delta(T)=\Delta_{\rm sc}$ was calculated from (\ref{eq: field velocity parameter}) \cite{thuneberg_2001}. The fit using these initial values was typically rather poor: below $T\approx0.55T_{\rm c}$ the calculated cf-peak height was exceeding the measured cf-peak height and the measured frequency shift was less than expected.

\begin{figure}[t]
	\begin{center}
		\begin{picture}(300,250)
			\put(0,0) { \includegraphics[width=0.85\textwidth]{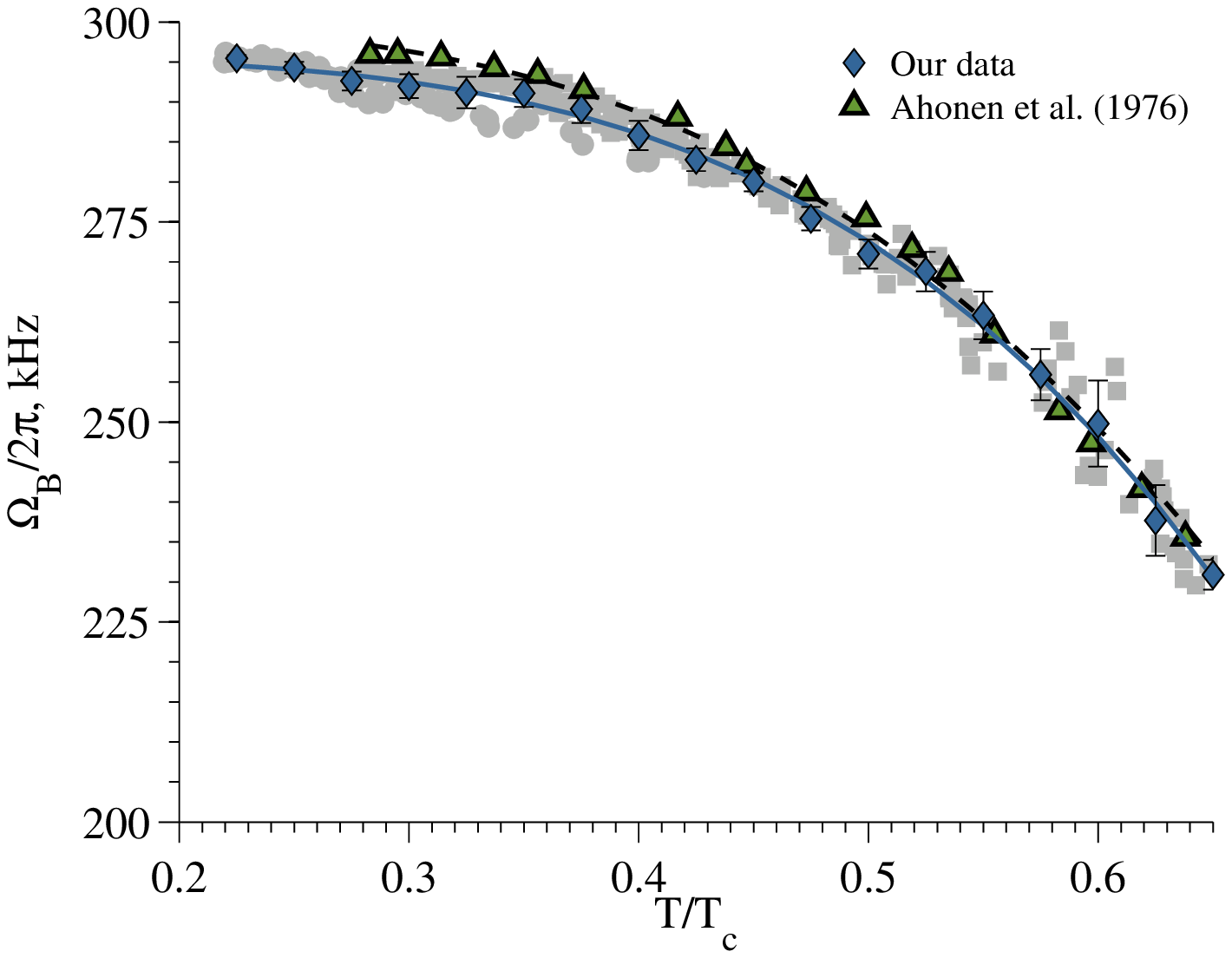} }
			\put(50,40){ \includegraphics[width=0.45\textwidth]{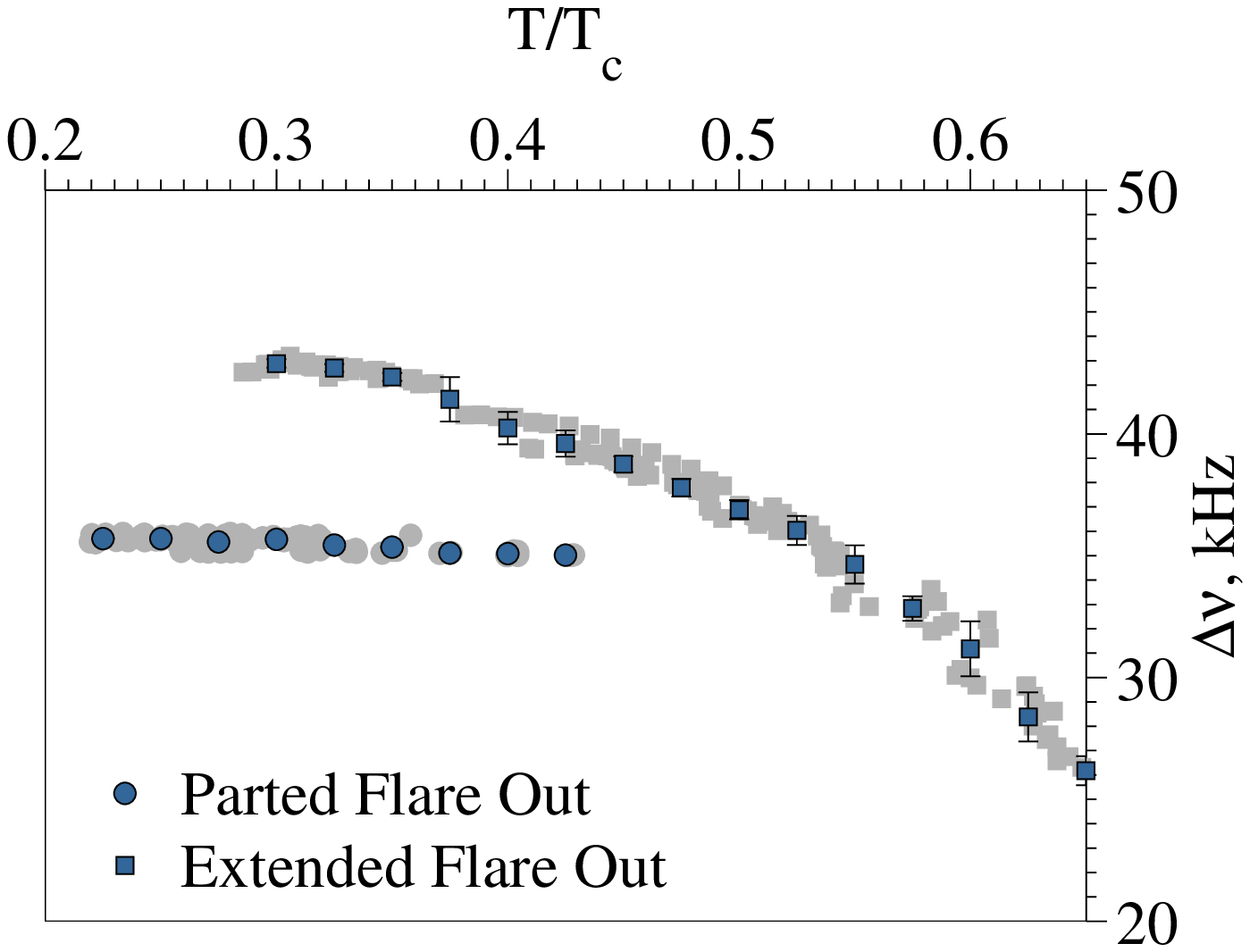} }
		\end{picture}
		\caption{\textbf{Longitudinal resonance frequency: $\Omega_{\rm B}(T,p)$} obtained from the fitted NMR line shape in the \textit{parted} and \textit{extended} flare out textures and a comparison with data from Ahonen \etal \cite{ahonen_1976}. The main panel shows the result of 321 runs (gray markers) and in binned form (blue circles). The inset shows the frequency shift from the Larmor frequency: in the \textit{parted} flare out texture the value of the magnetic field was obtained from the intersection of a tangent of the edge and the base-line; in the \textit{extended} flare out texture the magnetic field position of the $90^\circ$ peak was used. The frequency of the cw-NMR resonance circuit was $965.0\,$kHz and the fluid pressure $p=29\,$bar. Fits were done using the formula $(\Omega_{\rm B}/2\pi)^2=(1-\tilde{T}^4)(a-b\tilde{T}^4+c\tilde{T}^6)\cdot 10^{10}$ (reduced temperature $\tilde{T}=T/T_{\rm c}$, $\Omega_{\rm B}/2\pi$ in Hz); with parameter values $a=8.73312$, $b=13.32121$ and $c=1.51919$ for our data; and parameter values $a=9.00301$, $b=19.927$ and $c=15.3442$ for the data of Ahonen \etal (1976).}
		\label{fig:leggett frequency shift}
	\end{center}
\end{figure}

\begin{figure}[!tp]
	\begin{center}
		\begin{picture}(300,250)
			\put(10,10) { \includegraphics[width=0.85\textwidth]{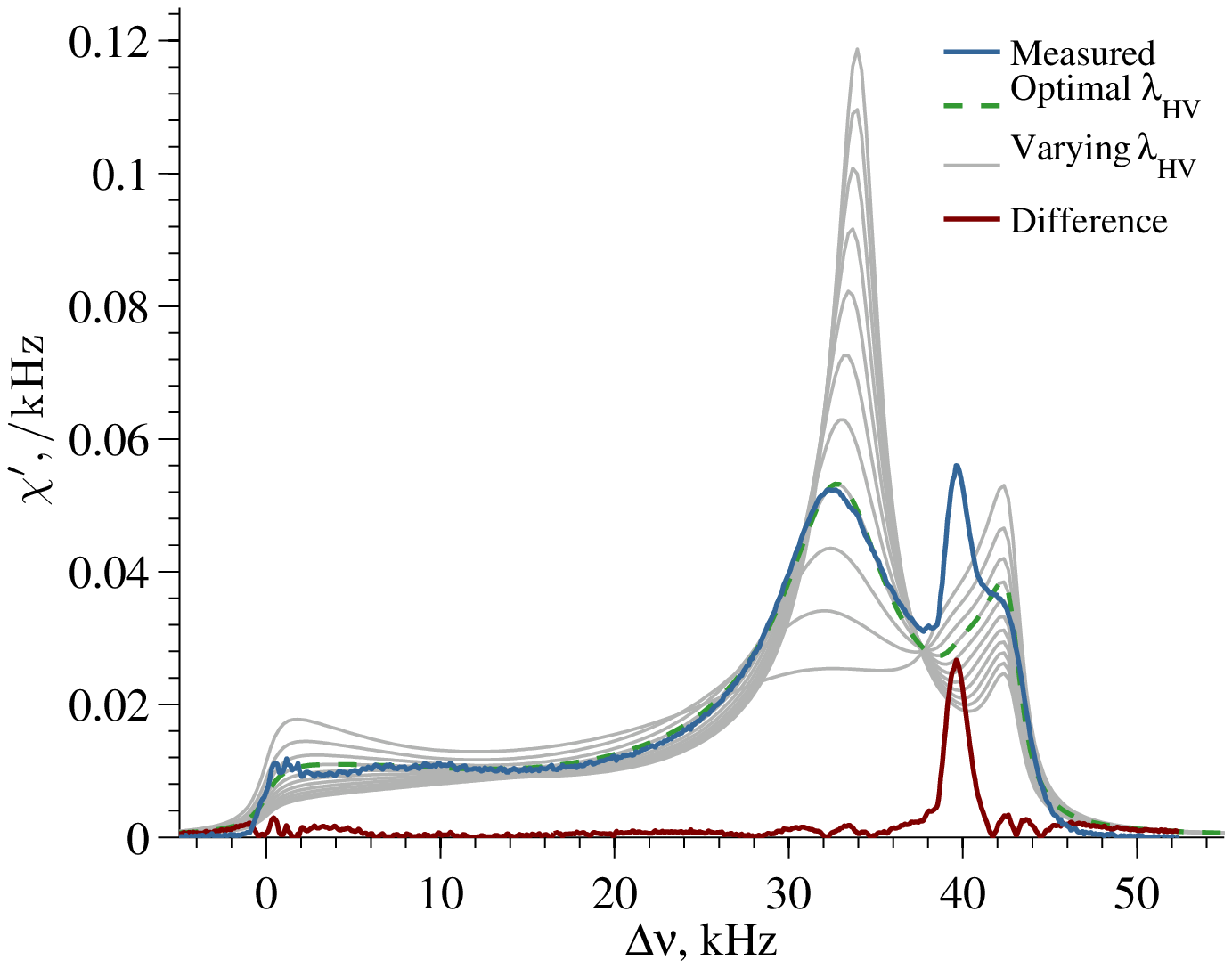} }
			\put(70,160){ \includegraphics[width=0.28\textwidth]{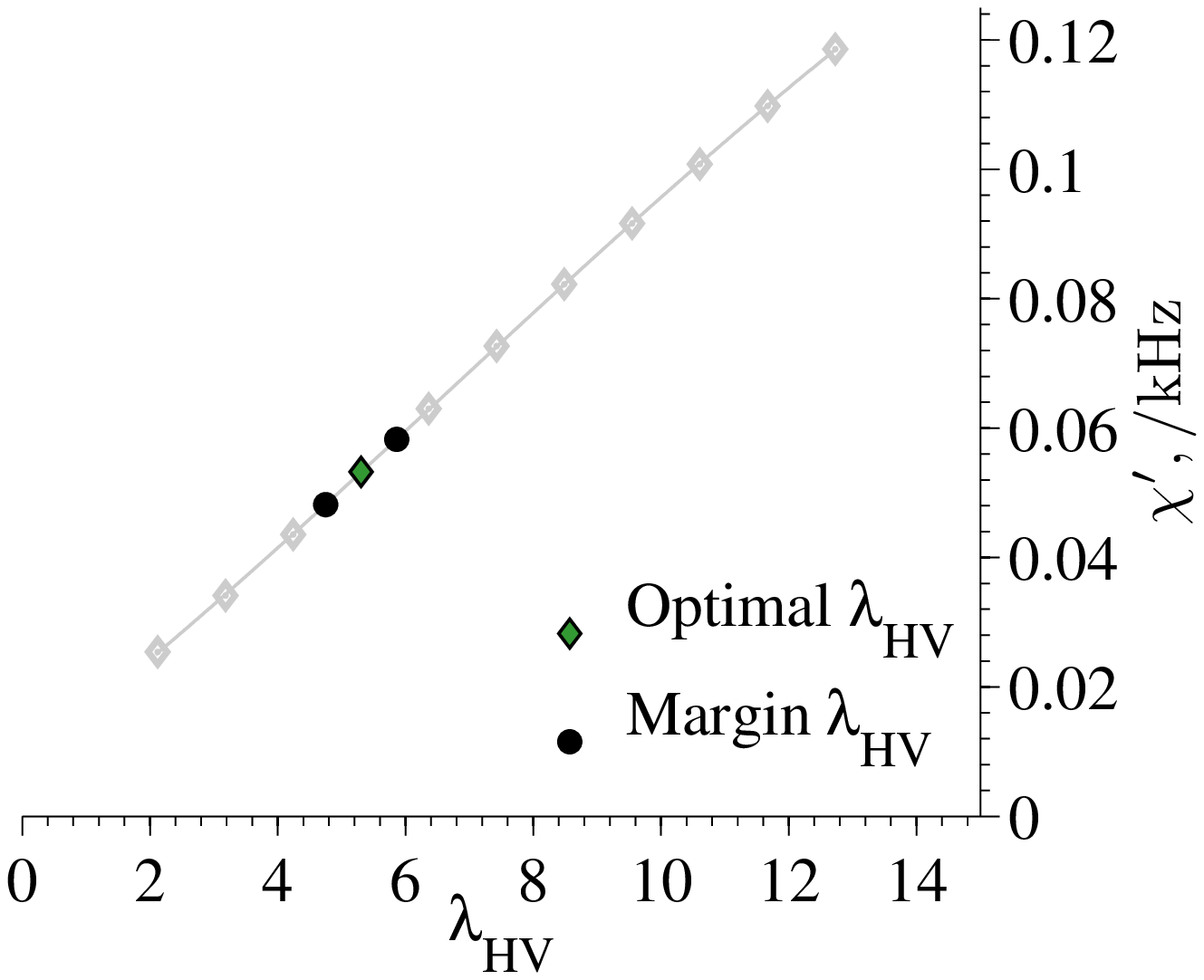} }
			\put(70,75) { \includegraphics[width=0.28\textwidth]{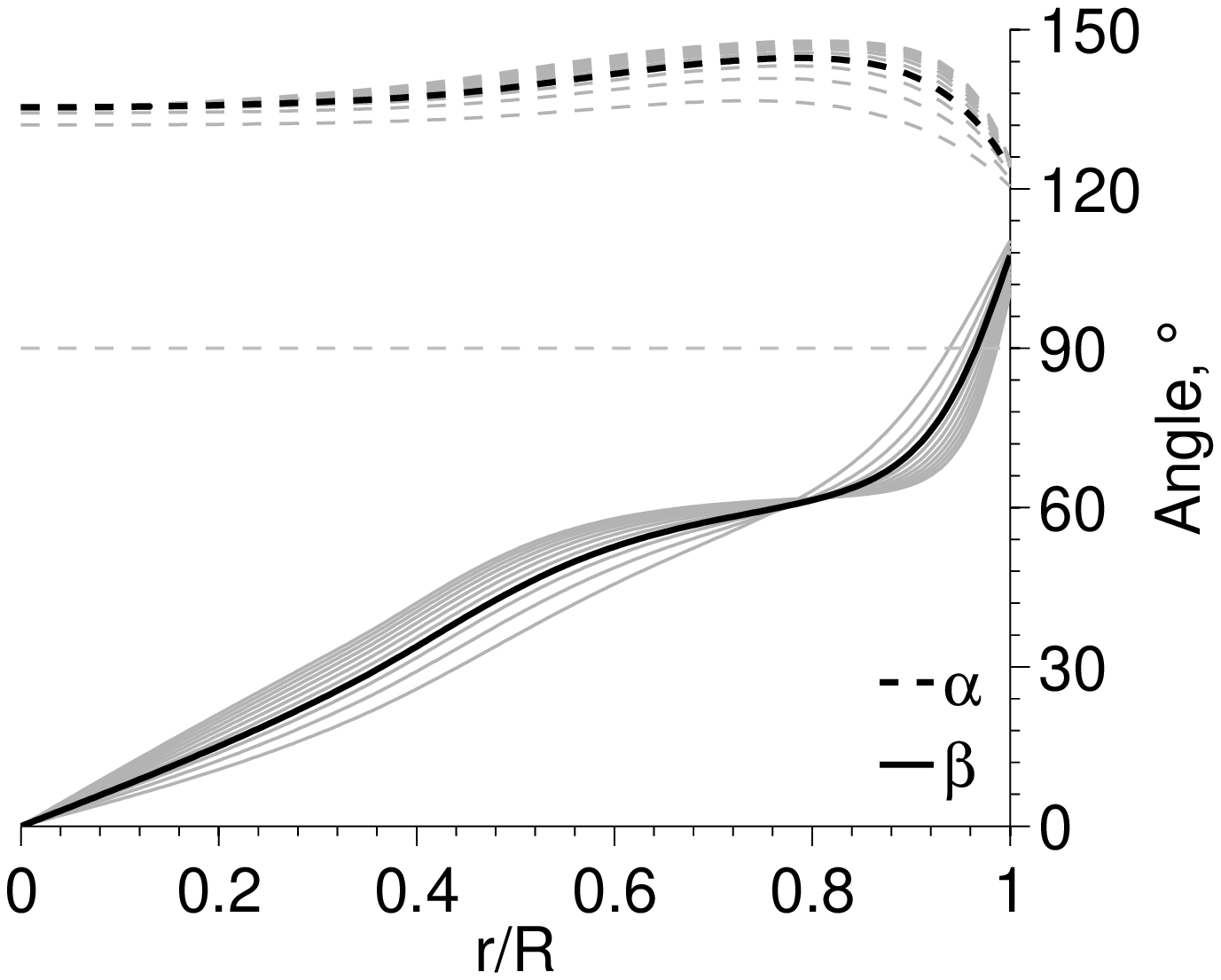} }
		\end{picture}
		\caption{\textbf{Uncertainty in the fitting procedure for the field velocity parameter $\lambda_{\rm HV}$ and isolation procedure of the surface peak}: the main panel shows the measured spectrum (blue) and the calculated spectrum (dashed green) using the optimal value for $\lambda_{\rm HV}$. The different line shapes drawn with gray curves are calculated by varying $\lambda_{\rm HV}$ around the optimal value (the bottom inset shows the corresponding $\hat{\vec{n}}$-vector textures), from which the \cf-peak dependence on $\lambda_{\rm HV}$ is determined, which is used for estimating the uncertainty in $\lambda_{\rm HV}$, see (\ref{eq: def uncertainty cfp}) and (\ref{eq: def uncertainty lambda_hv}). Taking the absolute difference between the measured spectrum (blue) and the calculated spectrum (dashed green) isolates the \textit{surface peak} (red). Note that the area under the surface peak is the main contributor to the uncertainty in $\lambda_{HV}$ as it directly influences the normalization of the measured spectrum. }
		\label{fig:fitting density anisotropy}
	\end{center}
\end{figure}

In the calculated spectra the frequency shift $\Delta\nu$ from the Larmor value is controlled by the temperature-dependent Leggett frequency $\Omega_{\rm B}(T)$ and the \cf-peak absorption by the field velocity parameter $\lambda_{\rm HV}$. The spread in the spectrum stems from the field inhomogeneity $\Delta H/H$, which influences directly the \cf-peak height. We adjusted the temperature, which solved in first approximation the deviation in the frequency shifts from the Larmor value. By manually adjusting the field inhomogeneity and/or the field velocity parameter $\lambda_{\rm HV}$ in the individual spectra did not give a unique solution for the best fit, since changes in these parameters are not orthogonal. As an illustration, see the cf-peak behavior for different values of $\lambda_{\rm HV}$ in \fig \ref{fig:fitting density anisotropy} and consider that changes in $\Delta H/H$ also affect the peak broadening and peak height in linear NMR: at high counterflow velocities the cf-peak height increases with increasing $\lambda_{\rm HV}$ as well as for decreasing values of $\Delta H/H$. However, using both the \textit{parted} and \textit{extended} flare out line shapes at the same temperature when the texture transition happens at $\Omega_{\rm c2}$ one can find a unique solution. Another method is to use the value of the critical velocity $\Omega_{\rm c1}$ for the determination of $\lambda_{\rm HV}$: higher values of $\lambda_{\rm HV}$ move $\Omega_{\rm c1}$ to a lower value. Subsequently a spectrum measured at high rotation velocity $\Omega$ is used for the  determination of the field inhomogeneity $\Delta H/H$. After the use of either method, a next iteration by correcting the temperature improves the fit further.

The results of fitting the Leggett longitudinal frequency $\Omega_{\rm B}$ as a function of the measured fork temperature are plotted in \fig \ref{fig:leggett frequency shift} and compared to data from Ahonen \etal \cite{ahonen_1976}. The temperature reading for our data is taken from the fork resonance width while that in \cite{ahonen_1976} was obtained by NMR measurement of fine platinum powder immersed in $^3$He-B. The inset shows the frequency shift from Larmor, $\Delta \nu = \nu - \nu_{\rm L}$, for both the \textit{parted} and \textit{extended} flare out textures: this result depends on the magnitude of the applied field $H$ or rf frequency, while the result for the main panel is general.

To obtain $\Omega_{\rm B}$ directly from the measured frequency shifts using equations (\ref{eqn: larmor frequency shift})...(\ref{eq: reduced freq shift}), one has to resort to a feature with known angle $\beta$. In the \textit{simple} flare out texture, this would be the edge with the highest shift from the Larmor value; in the \textit{extended} flare out texture the position of the $90^\circ$-peak can be used and in textures with high cf velocity the cf-peak position is at $\Delta\tilde{\nu}=0.8$. See for example \fig 1 in \cite{korhonen_1990}. Each method has its difficulty in determining the correct value, in particular when the position of the observed feature is on a slope. From numerical calculations we see that the observed $90^\circ$-peak is shifted due to the field inhomogeneity. The error in the frequency shift is estimated to be below 2\%, which is considered large. The error can be decreased by measuring spectra with a lower value of $\omega_{\rm L}$ such that the spectrum is wider, see (\ref{eqn: larmor frequency shift}), but this also reduces the cf peak sensitivity.

Even for the best fit there still exists a small discrepancy between the measured ($\chi'_{\rm m}$) and calculated ($\chi'_{\rm c}$) line shapes, which is expressed by the absolute value of the difference between the two line shapes normalized to the area of the measured spectra,
\begin{eqnarray}
	\Delta A = \frac{\int  | \chi'_{\rm m}(\nu) - \chi'_{\rm c}(\nu)|  d\nu}
	{\int \chi'_{\rm m}(\nu) d\nu } {\rm .}
	\label{eq: area between spectra}
\end{eqnarray}
We use this value for the uncertainty in $\lambda_{\rm HV}$ by calculating the spectra with different values of $\lambda_{\rm HV}$ around the optimum, which gives the relationship between the cf-peak height $\chi'_{\rm cfp}$ and $\lambda_{\rm HV}$, \ie $\lambda_{\rm HV}(\chi'_{\rm cfp})$. By defining the uncertainty in the cf-peak height $\chi'_{\rm cfp\pm}$, the uncertainty in $\lambda_{\rm HV}$ is also found
\begin{eqnarray}
	\chi'_{{\rm cfp} \pm} & = & \chi'_{\rm cfp} \cdot ( 1\pm \Delta A) {\rm ,}
	\label{eq: def uncertainty cfp} \\
	\lambda_{\rm HV\pm}    & = & \lambda_{\rm HV} (\chi'_{\rm cfp \pm} ) {\rm .}
	\label{eq: def uncertainty lambda_hv}
\end{eqnarray}
The procedure is demonstrated in \fig \ref{fig:fitting density anisotropy}. In that figure, the optimal value of $\lambda_{\rm HV}$ for fitting the calculation and the measurement is $\lambda_{\rm HV} = 5.3 {\rm kg/m^3T^2}$, which gives a cf-peak height $\chi'_{\rm cfp}=0.053\,{\rm kHz^{-1}}$. Increasing the field velocity parameter to $\lambda_{\rm HV} = 6.3\,{\rm kg/m^3T^2}$ gives the first gray line above the optimum calculated spectrum and the cf-peak increases to $\chi'_{\rm cfp}=0.063\,{\rm kHz^{-1}}$.  At the upper bound value of $\lambda_{\rm HV} = 5.8\,{\rm kg/m^3T^2}$, the absolute value of the difference between the line shapes, $\Delta A$, is fully attributed to the uncertainty in $\lambda_{\rm HV}$. In this figure the relation between $\lambda_{\rm HV}$ and $\chi'_{\rm cfp}$ is approximately linear, but this is not generally the case. In particular, when the total area under of the NMR line shape is almost completely under the cf-peak (high counterflow), the dependence of $\chi'_{\rm cfp}$ on $\lambda_{\rm HV}$ becomes almost constant and this increases the uncertainty $\lambda_{\rm HV \pm}$.

\begin{figure}[!tp]
	\begin{center}
		\includegraphics[width=0.9\textwidth]{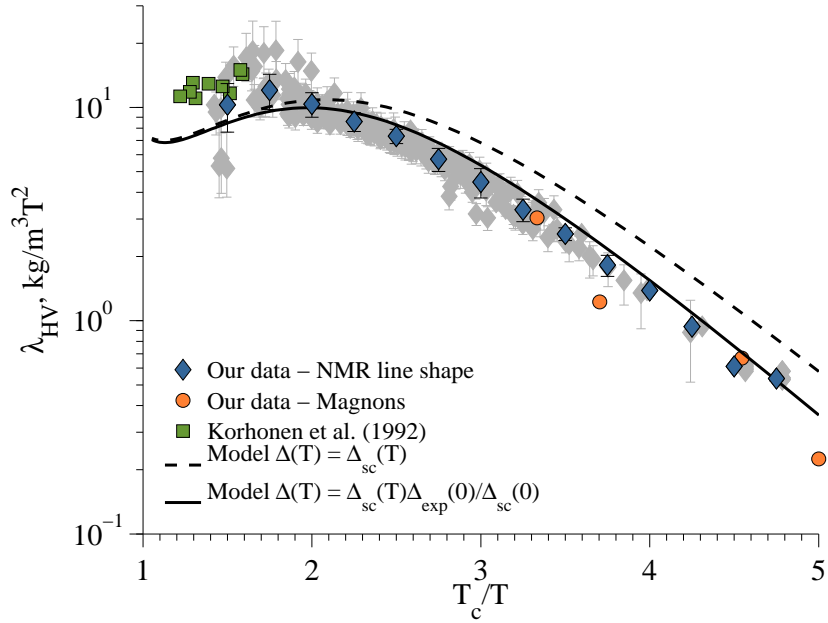}
		\caption{\textbf{Field velocity parameter $\lambda_{\rm HV}$ as a function of inverse temperature}. Our data using NMR line shape measurements (gray markers, 350 individual NMR spectra), $\lambda_{\rm HV}$ was obtained using the fitting procedure as described in \fig \ref{fig:fitting density anisotropy}. The error-bars indicate the boundaries of $\lambda_{\rm HV}$ as determined by the area mismatch, see (\ref{eq: def uncertainty lambda_hv}). Blue markers show averages within the inverse-temperature bins; on these markers the error-bars indicate $\pm 1$ standard deviation. In the low temperature regime, data is shown where $\lambda_{\rm HV}$ is obtained from fitting $\lambda_{\rm HV}$ in experiments with magnons in the potential well. In the high temperature regime data is shown from Korhonen \etal \cite{korhonen_1992}. Solid and dashed lines represent the model of the field velocity parameter $\lambda_{\rm HV}$ (\ref{eq: field velocity parameter}) with the energy gaps $\Delta_{\rm sc}$ (\ref{eq: gap sc}) and $\Delta_{\rm scaled}$ (\ref{eq: gap scaled}). }
		\label{fig:lambda_hv vs inverse temperature}
	\end{center}
\end{figure}

\fig \ref{fig:lambda_hv vs inverse temperature} shows the measured field velocity parameter $\lambda_{\rm HV}$ as a function of inverse temperature as measured with the forks. Gray markers are the results of individual measurements and blue markers are averages of this data within a bin of inverse temperature width $\Delta(T_{\rm c}/T)=0.25$. In the high temperature regime, our data compares well with those of Korhonen \etal \cite{korhonen_1992}. In the low temperature regime additionally data is included from our measurements of $\lambda_{\rm HV}$ using coherent precession NMR mode. This mode corresponds to the Bose-Einstein condensation of magnons on the ground level of a 3-dimensional potential well, which is created by the order parameter texture and applied profile of the static NMR field \cite{qballprl}. The position of the level, which can be measured with NMR, is a sensitive probe for the dependence $\beta(r)$ close to the axis of the sample (while the linear NMR response, explained above, is more sensitive to the $\beta(r)$ behavior at larger radii). The measured slope $\frac{d\beta}{dr}|_{r=0}$ can be connected with the relevant textural parameters via numerical calculations of the texture. The technique is explained in \cite{lambdameas}, where measurements are done in the equilibrium vortex state and the vortex textural parameter $\lambda$ is extracted. In this work we have performed analogous measurements in the vortex-free state using the upper spectrometer in the \textit{2005 setup} and fitted the magnon level positions using $\lambda_{\rm HV}$ as a fitting parameter. The results obtained using this non-linear NMR mode agree well with the measurements using linear NMR response.

Calculated values for $\lambda_{\rm HV}$ (see (\ref{eq: field velocity parameter})) are shown in \fig \ref{fig:lambda_hv vs inverse temperature} for the energy gap with strong-coupling correction, $\Delta_{sc}(T)$ (dashed line) \cite{thuneberg_2001}, sometimes also referred to as weak-coupling plus ($\Delta_{\rm wc+}$), and for the scaled energy gap $\Delta_{\rm scaled}(T)$. For the definitions of these energy gaps, see (\ref{eq: gap sc}) and (\ref{eq: gap scaled}). Measurements at pressures $p<19.4\,$bar by Davis \etal \cite{davis_2008} using transverse and longitudinal sound in $^3$He-B suggest the value of the energy gap $\Delta_{\rm sc}$.

The temperature in our data is obtained from the resonance frequency width $\Delta\omega$ of the fork by sweeping the fork around its resonance frequency. The resonance width at high temperatures ($T>0.34T_{\rm}$) is then calibrated against the MCT as a function of temperature. In the low temperature regime the dependence on temperature is extrapolated as an exponential dependence on the energy gap: $\Delta\omega \propto \exp(-\Delta/T)$. All temperature measurements of our data in this paper use the energy gap value $\Delta_{\rm exp}(0)=1.97\,k_{\rm B}T_{\rm c}$. When we compare the experimental data  with predictions of the theory using alternative energy gaps (\ref{eq: gap wc})...(\ref{eq: gap scaled}), we have to rescale the temperature measurement of the experiments, \ie use the energy gap $\Delta$ consistently. When we rescale the temperature of our data in \fig \ref{fig:lambda_hv vs inverse temperature} using the strong-coupling corrected gap $\Delta_{\rm sc}$, data points above $T_{\rm c}/T\approx3$ move to higher inverse temperature with maximum shift at $T_{\rm c}/T=5$. However, the temperature shift is minimal and the measured $\lambda_{\rm HV}$ has still a better correspondence using the scaled energy gap  with the low temperature limiting value $\Delta_{\rm exp}(0)=1.97\,k_{\rm B}T_{\rm c}$.

\section{Surface Spin Resonance}
\begin{figure}[t]
	\begin{center}
		\begin{picture}(300,250)
			\put(0,0) { \includegraphics[width=0.85\textwidth]{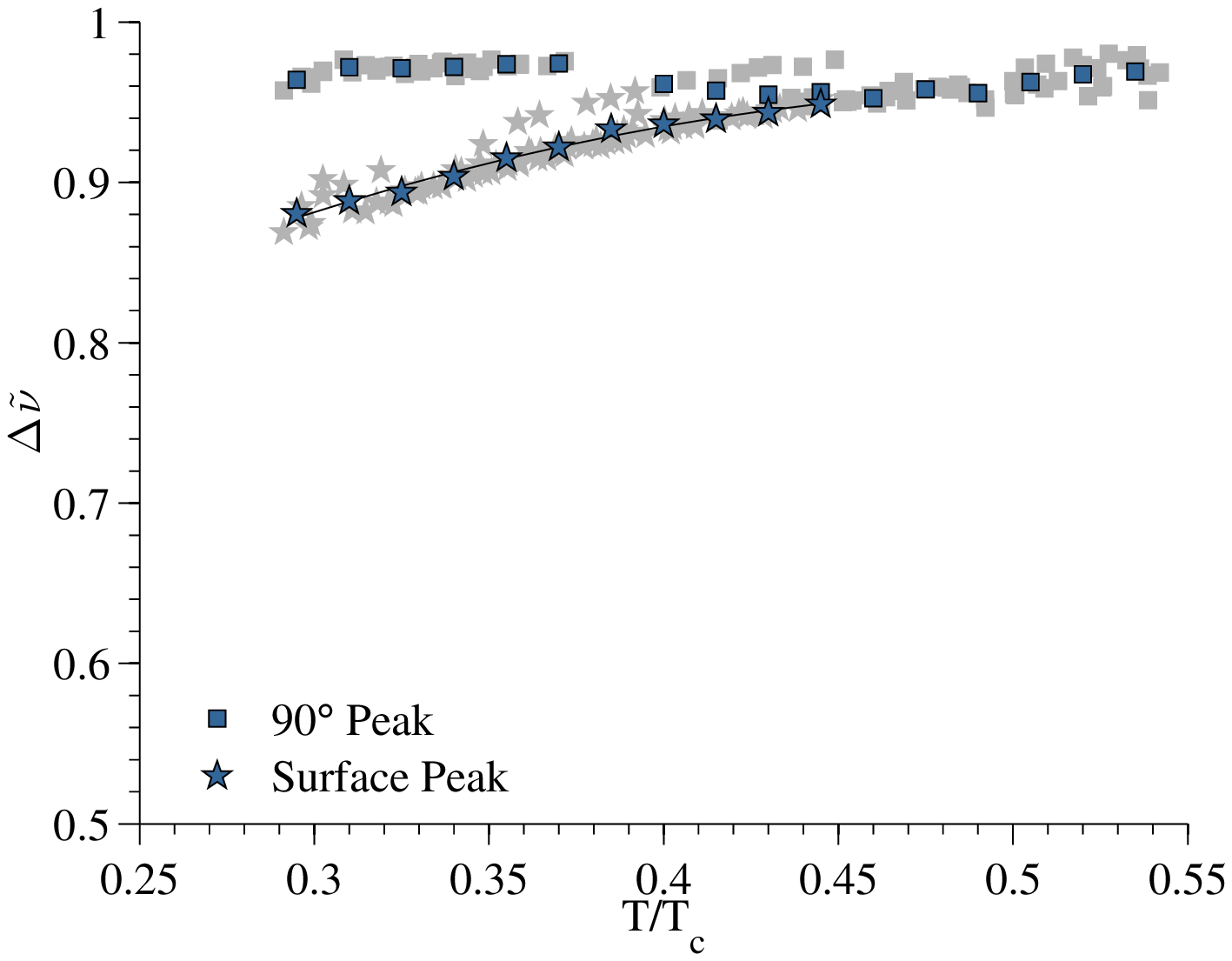} }
			\put(128,40){ \includegraphics[width=0.50\textwidth]{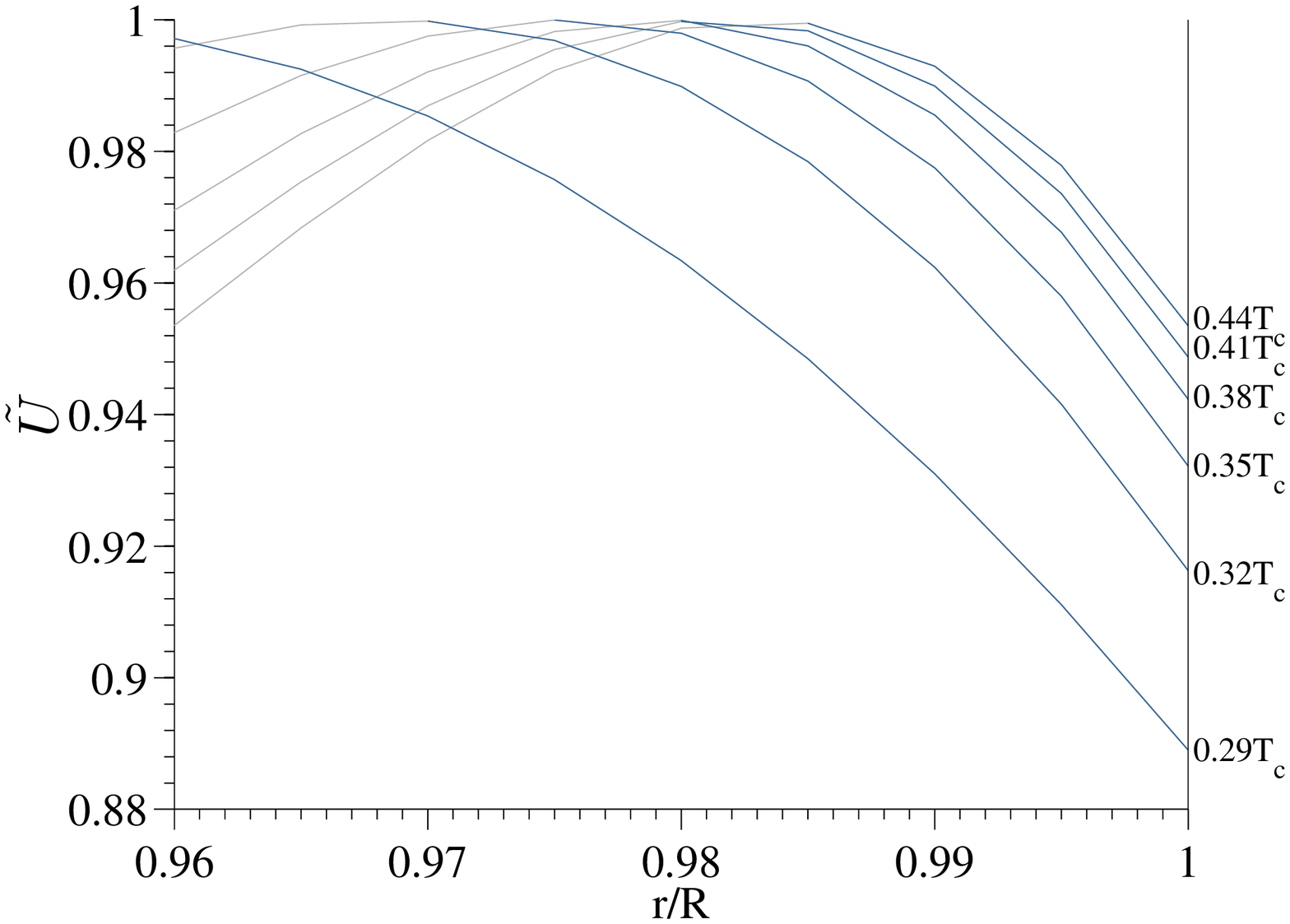} }
		\end{picture}
		\caption{\textbf{Surface peak measurements}: the main panel shows the reduced frequency shift $\Delta\tilde{\nu}=\sin^2\beta$ of the $90^\circ$-peak (squares) and the surface peak (stars) as a function of reduced temperature. Gray markers are individual measurements while blue markers are averages in temperature bins of width $\Delta T=0.015\,T_{\rm c}$. The inset shows the reduced textural energy $\tilde{U}(r)=\sin^2\beta(r)$ as a function of reduced radius at different temperatures in a cylinder of radius $R=3\,$mm. The curves illustrate the potential well at the cylindrical surface, which is formed between the location from where the $90^\circ$-peak originates and the surface boundary at $r=R$.
		}
		\label{fig: surface spin resonance}
	\end{center} 
\end{figure} 

In our measurements of the \textit{extended} flare out texture at temperatures $T < 0.45\,T_{\rm c}$, three absorption peaks have been observed in the NMR spectra (see \fig \ref{fig:textures and nmr line shapes}). The origin of the \cf-peak and the $90^\circ$-peak is identified in terms of the calculated texture and the local NMR response. The \cf-peak results from a large region of the sample where the angle $\beta$ is (approximately) constant and the $90^\circ$-peak from the maximum reduced frequency shift when $\beta$ crosses $90^\circ$, \ie $\Delta\tilde{\nu}=\sin^2(90^\circ)=1$. In \fig \ref{fig:fitting density anisotropy} the \textit{absolute} value of the difference (red line) is taken from the measured and calculated spectra which isolates a third peak. We call it the \textit{surface peak} based on the interpretation described below, which is an extension of one of the brief suggestions by Salomaa \cite{salomaa_1990}. 
A depiction of the peak appears also in later publications, see \fig 1 in \cite{korhonen_1990} and \fig 4 in \cite{finne_2004}.

Since the new peak is only observed in the \textit{extended} flare out texture, we suggest that it is associated with the particular $\hat{\vec{n}}(\vec{r})$ distribution close to the cylinder wall. In \fig \ref{fig: surface spin resonance} the reduced frequency shift $\Delta\tilde{\nu}$ of the surface peak is plotted against the reduced temperature $T/T_{\rm c}$. The result shows that the shift from the Larmor value decreases with decreasing temperature. The figure also contains the reduced frequency shift of the $90^\circ$-peak. When the separation of the surface peak from the $90^\circ$ peak is small, it becomes hard to distinguish these two peaks due to their overlap. The inset of \fig \ref{fig: surface spin resonance} shows the reduced textural energy $\tilde{U}(r) = \sin^2\beta(r)$ as a function of reduced radius at different temperatures. The decreasing energy at large radii creates a potential well for magnons between $\beta(r)=90^\circ$ ($\Delta\tilde{\nu}=\sin^2\beta=1$) and the cylindrical surface at $r=R$. We suggest that this potential well is the origin of the surface peak. The magnon eigenstates in the potential give rise to a coherent non-local spin-wave resonance \cite{bunkov_2010}. We suggest that the absorption at the frequency of the ground state forms the surface peak. With decreasing temperature the texture produces a potential well with lower energy, which corresponds qualitatively to a smaller frequency shift from the Larmor frequency.

\section{Conclusions}

In this work we have compared measured NMR line shapes of rotating superfluid $^3$He-B in a cylindrical environment with calculated line shapes using the Hydrostatic Theory of $^3$He-B. Quantitative agreement can be achieved by adjusting the field velocity parameter $\lambda_{\rm HV}$. The measured values of the parameter $\lambda_{\rm HV}$ show a deviation from the prediction for $\Delta(T)$ with strong-coupling correction,\ie $\Delta_{\rm sc}(0)=1.87\,k_{\rm B}T_{\rm c}$. A better correspondence between the measured and predicted values of $\lambda_{\rm HV}$ is established when $\Delta(T)$ is shifted toward a higher value $\Delta_{\rm exp}(0)=1.97\,k_{\rm B}T_{\rm c}$ as earlier measured by Todoshchenko \etal (2002).

Measurements on the susceptibility $\chi_{\rm B}/\chi_{\rm N}$ have been extended to lower temperatures ($T=0.2\,T_{\rm c}$). The low temperature limiting value of $\chi_{\rm B}/\chi_{\rm N}$
agrees with the currently accepted value of the anti-symmetric Fermi-liquid parameter $F^{\rm a}_0=-0.75$ at $29\,$bar liquid pressure.

By fitting the calculated and measured NMR spectra, we can determine the B-phase longitudinal resonance frequency $\Omega_{\rm B}(T,p)$, which shows a slight deviation below $T<0.5\,T_{\rm c}$ from the data measured by Ahonen \etal \cite{ahonen_1976}. Due to uncertainty in the measured pressure, we can not claim our result in the pressure dependent $\Omega_{\rm B}(T,p)$ is better. Since we here fit the whole spectrum which includes the broadening due to the field inhomogeneity, we believe that this procedure represents an improvement over the method used by Ahonen \etal

Taking the absolute difference between the measured and calculated spectra revealed in the \textit{extended} flare out texture below $0.45\,T_{\rm c}$ a new absorption peak, which we interpret as a spin-wave resonance in the potential well between the $90^\circ$-peak and the cylindrical surface. There is qualitative agreement between the reduced frequency shift of this peak and the shift in the potential well with decreasing temperature.

\newpage

\begin{acknowledgements}
The authors thank Juha Kopu for making his optimization program available. Erkki Thuneberg and Grigory Volovik are thanked for the theoretical discussions. Contributions to the measurements and the experimental setup by Antti Finne, Roman Solntsev and Rob Blaauwgeers are greatly appreciated. This work is supported in part by the Academy of Finland (Centers of Excellence Programme 2006-2011) and EU 7th Framework Programme (grant 228464 Microkelvin)
\end{acknowledgements}


\begin{thebibliography}{} 

\bibitem{kopu_2000}J. Kopu, R. Schanen, R. Blaauwgeers, V.B. Eltsov, M. Krusius, J.J. Ruohio and E.V. Thuneberg, J. Low Temp. Phys. \textbf{120}, 213 (2000)
 
%
\bibitem{hanninen_2009}R. H\"{a}nninen, V.B. Eltsov, A.P. Finne, R. de Graaf, J. Kopu, M. Krusius and R.E. Solntsev, J. Low Temp. Phys. \textbf{155}, 98 (2009)

%
\bibitem{solntsev_2007}R.E. Solntsev, R. de Graaf, V.B. Eltsov, R. H\"{a}nninen and M. Krusius, J. Low Temp. Phys. \textbf{148}, 311 (2007)

\bibitem{de_graaf_2008}R. de Graaf, R. H\"{a}nninen, T.V. Chagovets, V.B. Eltsov, M. Krusius and R.E. Solntsev, J. Low Temp. Phys. \textbf{153}, 197 (2008)

\bibitem{eltsov_2007} V.B. Eltsov, A.I. Golov, R. de Graaf, R. H\"{a}nninen, M. Krusius, V.S. L'vov, R.E. Solntsev,  Phys. Rev. Lett. \textbf{99}, 265301 (2007)

\bibitem{eltsov_2010}V.B. Eltsov, R. de Graaf, P.J. Heikkinen, J.J. Hosio, R. H\"{a}nninen, M. Krusius, and V.S. L'vov, Phys. Rev. Lett. \textbf{105}, 125301 (2010)

%
\bibitem{thuneberg_2001}E.V. Thuneberg, J. Low Temp. Phys. \textbf{122}, 657 (2001)

\bibitem{kopu_2007}J. Kopu, J. Low Temp. Phys. \textbf{146},47 (2007)

\bibitem{osheroff_1974}D.D. Osheroff and W.F. Brinkman, Phys. Rev. Lett. \textbf{32}, 584 (1974)

\bibitem{balian_1963} R. Balian and N.R. Werthamer, Phys. Rev. \textbf{131}, 1553 (1963)

\bibitem{vollhardt_1990}D. Vollhardt and P. W\"{o}lfle, The Superfluid Phases of Helium 3, 81, Taylor \& Francis, London (1990)


\bibitem{tn_2010}$http://www.netlib.org/opt/tn$

\bibitem{hakonen_1989}P.J. Hakonen, M. Krusius, M.M. Saloma, R.H. Salmelin, J.T Simola, A.D. Gongadze, G.E. Vachnadze and G.A. Kharadze, J. Low Temp. Phys. \textbf{76}, 225 (1989) 

%
\bibitem{salomaa_1990} M.M. Salomaa, J. Phys-Condens Matter \textbf{2}, 1325 (1990)

\bibitem{brinkman_1974}W.F. Brinkman, H. Smith, D.D. Osheroff and E.I. Blount, Phys. Rev. Lett. \textbf{33}, 624 (1974)

\bibitem{eltsov_2010_2}V.B. Eltsov, R. de Graaf, P.J. Heikkinen, J.J. Hosio, R. H\"{a}nninen and M. Krusius, J. Low Temp. Phys. \textbf{161}, 474 (2010)

\bibitem{greywall_1986} D.S. Greywall, Phys. Rev. B \textbf{33}, 7520 (1986)

\bibitem{bevan_1997}T.D.C. Bevan, A.J. Manninen, J.B. Cook, H. Alles, J.R. Hook and H.E. Hall, J. Low Temp. Phys. \textbf{109}, 423 (1997)

\bibitem{todoshchenko_2002}I.A. Todoshchenko, H. Alles, A. Babkin, A.Y. Parshin and V. Tsepelin, J. Low Temp. Phys. \textbf{126}, 1449 (2002)

\bibitem{dobbs_2000}E.R. Dobbs, Helium Three, 512, Oxford University Press, New York (2000)

\bibitem{bozler_1992}H.M. Bozler, P.D. Saundry, Inseob Hanh, S.T.P. Boyd and C.M. Gould, J. Low Temp. Phys. \textbf{89}, 5 (1992)

\bibitem{scholz_1981} H.N. Scholz, Ph.D. thesis, Ohio State University, 1981 (unpublished)

\bibitem{inseob_hahn_1998} Inseob Hahn, S.T.P. Boyd, H.M. Bozler and C.M. Gould, Phys. Rev Lett. \textbf{81}, 618, 1998

\bibitem{korhonen_1990}J.S. Korhonen, A.D. Gongadze, Z. Jan\'u, Y. Kondo, M. Krusius and E.V. Thuneberg, Phys. Rev. Lett. \textbf{65}, 1211 (1990)

%
\bibitem{korhonen_1992}J.S. Korhonen, Y.M. Bunkov, V.V. Dmitriev, Y. Kondo, M. Krusius, Y.M. Mukharskiy, \"{U}. Parts and E.V. Thuneberg, Phys. Rev. B \textbf{46}, 13983 (1992)

\bibitem{qballprl} Yu.M. Bunkov and G.E. Volovik, Phys. Rev. Lett. \textbf{98}, 265302 (2007)

\bibitem{lambdameas} V.B. Eltsov, R. de Graaf, M. Krusius and D.E. Zmeev, J. Low Temp. Phys. \textbf{161} (2011) [DOI: 10.1007/s10909-010-0285-1]. 

\bibitem{ahonen_1976}A.I. Ahonen, M. Krusius and M.A. Paalanen, J. Low Temp. Phys. \textbf{25}, 421 (1976)

\bibitem{osheroff_1972}D.D. Osheroff, W.J. Gully, R.C. Richardson and D.M. Lee, Phys. Rev. Lett. \textbf{29}, 920 (1972)

\bibitem{davis_2008}J.P. Davis, H. Choi, J. Pollanen and W.P. Halperin, J. Low Temp. Phys. \textbf{153}, 1 (2008)

\bibitem{finne_2004}A.P. Finne, S. Boldarev, V.B. Eltsov and M. Krusius, J. Low Temp. Phys. \textbf{136}, 249 (2004)

\bibitem{bunkov_2010} Yu.M. Bunkov, V.B. Eltsov, R. de Graaf, P.J. Heikkinen, J.J. Hosio, M. Krusius and G.E. Volovik, arXiv:1002.1674v1 [cond-mat.quant-gas], (2010)


		




\end{thebibliography}
\end{document}